\documentclass[12pt]{article}

\usepackage{latexsym}

\usepackage[font=small,labelfont=bf]{caption}

\usepackage{graphics}
\usepackage{graphicx}
\usepackage{epstopdf}
\usepackage{amssymb}
\usepackage{tabularx}
\usepackage{caption}
\usepackage{subcaption}
\usepackage{hyperref}
\usepackage{tabularx}
\usepackage{gensymb}
\usepackage{mathtools,amsmath,amssymb,mathrsfs,color,esint}
\usepackage{array}
\usepackage{makecell}
\usepackage{float}
\usepackage{dcolumn}

\begin{document}


\title {Observing naked singularities by the present and next-generation Event Horizon Telescope}

\author{
Valentin Deliyski$^{1}$\footnote{E-mail: \texttt{vodelijski@phys.uni-sofia.bg}}, \,
Galin Gyulchev$^{1}$\footnote{E-mail: \texttt{gyulchev@phys.uni-sofia.bg}},  \,
Petya Nedkova$^{1}$\footnote{E-mail: \texttt{pnedkova@phys.uni-sofia.bg}},
\\ Stoytcho Yazadjiev$^{1,2}$\footnote{E-mail: \texttt{yazad@phys.uni-sofia.bg}}\\ \\
 {\footnotesize${}^{1}$ Faculty of Physics, Sofia University,}\\
  {\footnotesize    5 James Bourchier Boulevard, Sofia~1164, Bulgaria } \\
    {\footnotesize${}^{2}$ Institute of Mathematics and Informatics,}\\
{\footnotesize Bulgarian Academy of Sciences, Acad. G. Bonchev 8, } \\
  {\footnotesize  Sofia 1113, Bulgaria}}
\date{}
\maketitle

\begin{abstract}
We consider the observational signatures of reflective naked singularities as seen by the current and next-generation Event Horizon Telescope (EHT). The reflective naked singularities lead to a distinctive morphology of their accretion disk images producing  a series of bright rings at the central part of the image. We explore the capacity of the present and near-future EHT arrays to detect this structure considering two particular naked singularity spacetimes and modeling the galactic target M87$^*$. We obtain that the 2017 EHT array is incapable of resolving the bright ring series. However, it detects an increased overall intensity of the central brightness depression reaching with an order of magnitude higher values than for the Kerr black hole. This metric can be used as a quantitative measure for the absence of an event horizon. The observations with the next-generation EHT at 230 GHz would reveal two orders of magnitude difference in the intensity of the central brightness depression between naked singularities and black holes. Introducing a second observational frequency at 345 GHz would already  resolve qualitative effects in the morphology  of the  disk image for naked singularities as certain bright spots become apparent at the center of the image.
\end{abstract}

\section{Introduction}

The Event Horizon Telescope provided recently the first images of the galactic targets M87$^*$ and SgrA$^*$ opening a new era of precision tests in the strong gravity regime \cite{EHT1}-\cite{EHT_2}. In the next decade its resolution is expected to increase substantially by expanding the telescope arrays, introducing a second observational frequency at 345 GHz, and adding  space-based telescopes. These improvements will allow obtaining more detailed information about the gravitational sources constraining their nature and the specifics of the accretion flow in their vicinity.

Revealing the physical nature of the compact objects at the center of the  galactic nuclei is an important goal of the strong gravity tests. It may refute the Kerr hypothesis by providing evidence for the presence of new fundamental fields \cite{Amarilla:2010}-\cite{Kocherlakota:2021}. Furthermore, it may confirm the existence of more exotic compact objects like boson stars, gravastars, wormholes and naked singularities \cite{Nedkova:2013}-\cite{Vincent:2020}. These compact objects arise naturally in the quantum gravity motivated alternative theories of gravity and their experimental detection can serve as an observational evidence for deviation from general relativity.

The detection of exotic compact objects by the Event Horizon Telescope relies on observing certain characteristic features  in their images which distinguish them from black holes. These signatures arise as phenomenological effects caused by the properties of the gravitational lensing in the corresponding spacetimes. An important issue is to confront further the theoretical predictions with the capacity of the current and near-future EHT arrays. Due to the limited resolution the expected features may not be distinguishable or may be encoded indirectly into certain metrics of the images. Such questions were addressed for regular horizonless spacetimes such as boson stars, wormholes and quantum gravity motivated regular black holes in \cite{Vincent:2020}-\cite{Eichhorn:2023}.

The aim of this paper is to study the observational appearance of certain types of naked singularities as seen by the current and next-generation Event Horizon Telescope (ngEHT). Naked singularities which possess a photon sphere frequently mimic the phenomenological behavior of black holes \cite{Nedkova:2019}. However, there exist classes of naked singularities which possess screening properties with respect to null geodesics. In these spacetimes in-falling photon trajectories cannot access the curvature singularity since they are reflected by the gravitational field in its vicinity and scatter away to infinity.

In our previous work we studied the Janis-Newman-Winicour and the Gauss-Bonnet naked singularities as examples of such spacetimes \cite{Nedkova:2020}-\cite{Nedkova:2021}. These geometries are characterized by rather different lensing properties since the former possesses no photon sphere, while the latter contains a stable and an unstable light ring. However, due to the reflective behavior of the gravitational potential in the vicinity of the singularity they share similar phenomenological effects in their accretion disk images. Considering the Novikov-Thorne thin disk model we observed that the accretion disk images in these spacetimes do not possess a pronounced brightness depression in their center which is characteristic for black holes. Instead, they lead to the formation of a series of central bright rings providing  a qualitatively different signature in the disk image morphology \cite{Nedkova:2020}-\cite{Nedkova:2021}. In addition, the linear polarization of the emission from the central rings possesses a characteristic twist of the polarization direction around the ring which further distinguishes the reflective naked singularities from black holes \cite{Nedkova:2023}.

The next question which we address in our work is how well we can observe the central ring structure with the current capacity of the Event Horizon Telescope. We adopt physical settings which are relevant for the galactic source M87$^*$ and using a radiatively inefficient accretion flow (RIAF) model we simulate the accretion disk images which should be observed theoretically in the Janis-Newman-Winicour and Gauss-Bonnet spacetimes by a distant observer. Then, we take into account the limitations of the Event Horizon Telescope arrays and obtain the reconstructed images as they would appear in reality if the compact object at the center of  M87$^*$ is a naked singularity. We use the ehtim toolkit \cite{Chael:2018} with the 2017 and 2022 EHT arrays and a tentative next-generation EHT array which observes both at 230 GHz and 345 GHz. We further analyse the geometry of the accretion disk images performing a fitting procedure to a Gaussian ring template and estimating the ring diameter and its width. As a measure for the central brightness depression we adopt the ratio of the minimum image intensity at its center and the maximum intensity around the ring. This metric can quantify the deviation from black holes in the cases when the central ring structure is not resolved. The introduction of a second frequency leads to a significant observational improvement since a certain distribution of bright spots becomes apparent at the central region of the image. In this case the deviation from black holes is quantified by the maximum intensity of the bright spots normalized to the maximum intensity of the disk image.

The paper is organized as follows. In the next section we briefly review the Janis-Newman-Winicour and the Gauss-Bonnet naked singularities and the phenomenology of the their disk images using the Novikov-Thorne thin disk model. In section 3 we consider a phenomenological radiatively inefficient accretion flow (RIAF) model which describes more realistically the physical conditions in M87$^*$. In section 4 we obtain the theoretical images of the RIAF disks around the JNW and Gauss-Bonnet naked singularities performing a  numerical ray-tracing. We further reconstruct the realistic images as they would be seen by the 2017 EHT array using the ehtim toolkit. Finally, we investigate the geometry of the disk images by performing a fit with a ring template by means of the VIDA package. In section 5 we study the observational properties of the accretion disks around naked singularities  as seen by the upgraded next-generation EHT arrays. In the last section we present our conclusions.

\section{Reflective naked singularities}

We consider a class of naked singularities which possess the property of reflecting null geodesics so that they cannot access the curvature singularity. In these spacetimes the gravitational field becomes effectively repulsive for photon trajectories in the vicinity of the curvature singularity building a potential barrier which prevents them from reaching it.

As examples for such spacetimes we can consider the Janis-Newman-Winicour solution \cite{Fisher:1948} - \cite{Virbhadra:1997}

\begin{equation}
ds^2 = -\left(1-\frac{2M}{\gamma r}\right)^{\gamma} dt^2
      +\left(1-\frac{2M}{\gamma r}\right)^{-\gamma} dr^2
      + \left(1-\frac{2M}{\gamma r}\right)^{1-\gamma} r^2\left(d\theta^2 + \sin^2\theta d\phi^2\right),
\end{equation}
and the Einstein-Gauss-Bonnet naked singularity \cite{Cai:2010} - \cite{Tomozawa:2011}

\begin{eqnarray}
&&ds^2 = -f(r)dt^2 + \frac{1}{f(r)}dr^2 + r^2(d\theta^2 + \sin^2\theta \phi^2), \nonumber \\[2mm]
&&f(r) = 1 + \frac{r^2}{2\gamma M^2}\left(1-\sqrt{1+ \frac{8\gamma M^3}{r^3}}\right), \nonumber
\end{eqnarray}
for certain ranges of their parameters. Both solutions describe static spherically symmetric spacetimes characterized by a single parameter $\gamma$. In the case of the JNW naked singularity $\gamma$ is associated with the scalar charge of the solution, while for the Gauss-Bonnet solution it represents the coupling constant with the Gauss-Bonnet invariant.

For static spherically symmetric spacetimes we can analyse the qualitative behavior of the null geodesics by introducing the effective potential in the equatorial plane

\begin{eqnarray}
V_{eff}= L^2\frac{g_{tt}}{g_{\phi\phi}},
\end{eqnarray}
where $g_{\mu\nu}$ are the metric functions and $L$ is the specific angular momentum on the geodesic. Its functional form completely determines the possible types of photon trajectories. Thus, considering a scalar field parameter in the range $0<\gamma<0.5$ we obtain for the JNW singularity an effective potential which possesses no local extrema and diverges in the vicinity of the singularity (see Fig. 1 in \cite{Nedkova:2020}). This class of spacetimes contain no photon sphere and all the in-falling null geodesics scatter away to infinity.

On the other hand, if we analyse the Gauss-Bonnet naked singularities with $1<\gamma<3\sqrt{3}/4$ we obtain an effective potential with a local minimum and a local maximum which diverges again in the vicinity of the singularity (see Fig. 1 in \cite{Nedkova:2021}). This class of  spacetimes possess a stable and an unstable photon ring, thus allowing for more complicated dynamics of the photon trajectories. However, due to the presence of a potential barrier in the vicinity of the singularity all the in-falling null geodesics still get reflected and scatter away to infinity.

\begin{figure} [t]
\centering
\begin{tabular}{ cc}
           \includegraphics[width=0.9\textwidth]{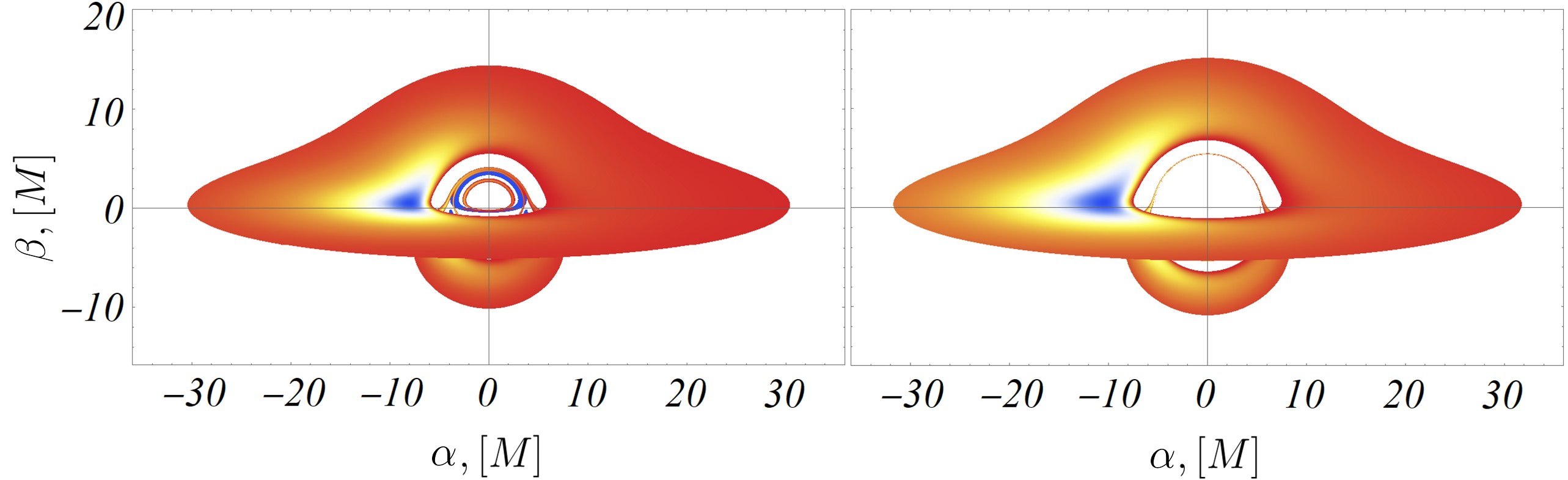}
                      \vspace{0.5cm}\includegraphics[width=0.068\textwidth]{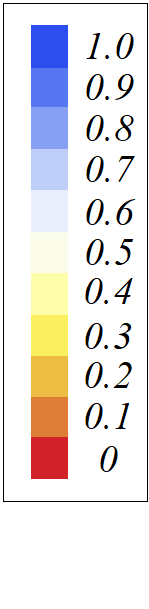}
        \end{tabular}
\caption{Image of the thin accretion disk around the Janis-Newman-Winicour singularity with scalar field parameter $\gamma=0.48$ (left panel) \cite{Nedkova:2020}. The Schwarzschild black hole is presented for comparison in the right panel.  The observer is located at $r=5000M$, while the inclination angle is $i=80^\circ$. }
\label{fig:JNW}
\end{figure}

\begin{figure}
\includegraphics[width=\textwidth]{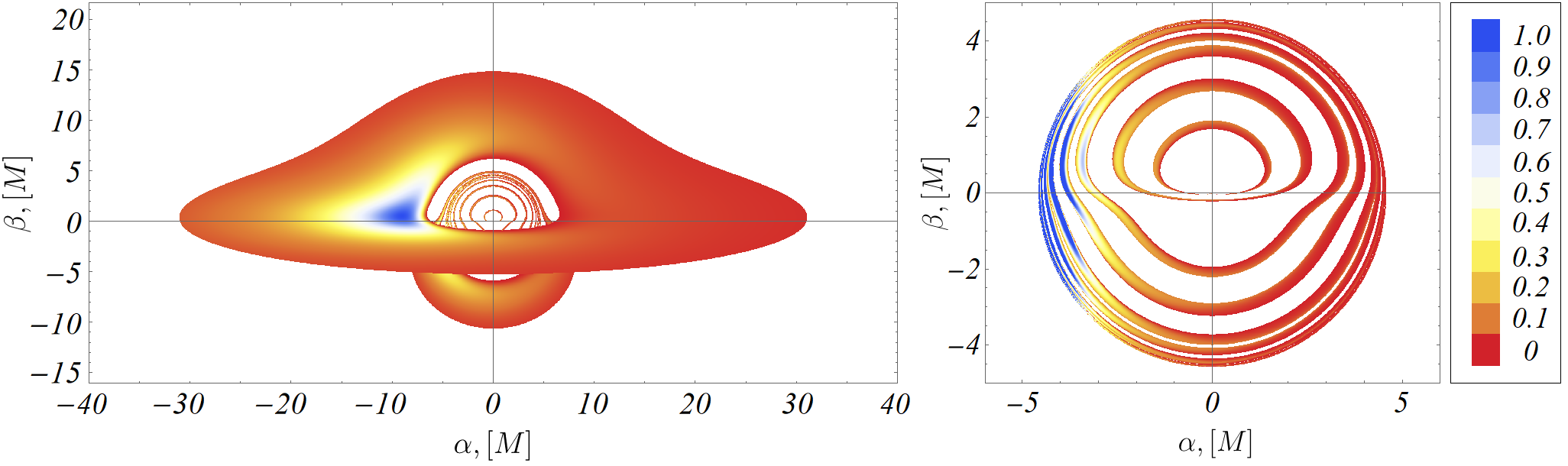}
\caption{Image of the thin accretion disk around the Gauss-Bonnet naked singularity with coupling constant $\gamma=1.15$ \cite{Nedkova:2021}. The observer is located at $r=5000M$, while the inclination angle is $i=80^\circ$.}
\label{fig:GB}
\end{figure}

The lensing properties of the compact objects influence directly their imaging in the astrophysical experiments, in particular the observable images of the accretion disks around them. The described reflective behavior for the null geodesics introduces  observational signatures in the accretion disk images which are absent for black holes. The apparent shape of the accretion disk around black holes is characterized by a central brightness depression which corresponds to the black hole shadow. The reflective naked singularities lead instead to a series of bright rings in the central part of the disk image which correspond to  double images of the disk of different order (see Figs. $\ref{fig:JNW}$-$\ref{fig:GB}$). This effect was first discovered and explained by studying thin  disks within the Novikov-Thorne model. In this case it was demonstrated that the radiation from the central rings is substantial and it may even represent the maximum radiation from the disk image.

A viable issue is whether the apparent central ring structure will be preserved for more general models of accretion. The appearance of the ring structure is caused by a fundamental property of the spacetime geometry. Therefore, the characteristic morphology of the disk images is expected to be preserved when varying the accretion model. However, the intensity of the emission from the central rings will probably be modified and the resolution of the Event Horizon Telescope may be insufficient to distinguish the central ring structure. We explore these issues in the next sections  considering  the galactic target M87$^*$ and the properties of the current and near-future EHT arrays.

\section{Accretion disk model}

We consider an analytical model describing a slim radiatively inefficient accretion disk (RIAF) which is qualitatively consistent with the GRMHD simulations of magnetically arrested disk (MAD) models \cite{Narayan:2003},\cite{McKinney:2012}, \cite{EHT6}. The disk emission is produced by  a synchrotron radiation from thermal and non-thermal electrons. Following \cite{Broederick:2006} the density of  both electron populations can be represented as a power law in the radial direction and a Gaussian profile in the vertical direction. In order to be able to fine-tune the location of the emission region we further introduce an exponential term serving as a radial cut-off of the radiation in the vicinity of the compact object.  Thus, we consider the following density profile \cite{Broederick:2022}-\cite{Gold:2020}

\begin{align}
\label{disk_model}
	n_e(r, z) =
	n_0\,
	\left(\frac{r}{r_0}\right)^{-2}e^{-\frac{z^2}{2\,(\alpha\rho)^2}}\,
	\begin{cases}
		e^{-\frac{(r-r_0)^2}{r_{sc}^2}}\;, & 0<r<r_0 \\
	1\;, & r>r_0
	\end{cases}\;,
\end{align}
while the electron temperature is

\begin{eqnarray}
T_e(r)&=T_0\left(\frac{r}{r_0}\right)^{-1}.
\end{eqnarray}

We denote by $\rho$ and $z$ the cylindrical coordinates $\rho = r\sin\theta$, $z=r\cos\theta$ and the parameter $\alpha$ determines the opening angle of the disk. If we model the disk surface as a cone, we have the relation $\alpha = \tan\theta_0$, where $\theta_0$ is the opening angle of the cone measured with respect to the equatorial plane (see \cite{Vincent:2020}). In order to obtain a slim disk we set $\alpha =0.1$. The parameter $r_0$ corresponds to the inner cut-off radius of the disk emission while $r_{sc}$ defines the sharpness of the radiation profile in the vicinity of the compact object. The cut-off radius $r_0$ primarily determines the location of the maximum radiation in the observable disk image, and hence its diameter according to the image metrics defined in \cite{EHT4}. Therefore, we choose its value for every compact object in such a way that the angular size of the disk image is compatible with the observational data for M87$^*$. The sharpness parameter is fixed to  $r_{sc} =0.4$M throughout the paper.

The parameters $n_0$ and $T_0$ correspond to the equatorial electron number density and temperature at the cut-off radius, respectively,  and their values determine the observable flux of the disk. Since we have two parameters constrained by a single observable we can fix the number density $n_0$ to a reasonable value which agrees with the previous results in the literature and choose $T_0$ in each simulation in such a way that the flux is compatible with the observational data for M87$^*$. Thus, we fix $n_0=5\times 10^5\, \text{cm}^{-3}$ (see e.g. \cite{EHT5}) and fine-tune $T_0$ for each compact object so that we obtain flux in the range $0.5-0.6$ Jy.

In order to calculate the synchrotron radiation we assume a uniformly magnetized  disk. We consider the magnetization parameter

\begin{align}
    \sigma=\frac{B^2}{4\pi m_p c^2 n_e},
\end{align}
where $B$ is the magnetic field magnitude, $m_p$ is the proton mass, and $c$ is the speed of light. The magnetization parameter is fixed to $\sigma =0.01$ in the paper following previous work \cite{Vincent:2020}, \cite{Vincent:2022} as its value influences insignificantly our results.

The radiation from the disk is computed integrating the radiative transport equation \cite{Bronzwaer:2018}
\begin{eqnarray}
\frac{d(I_\nu/\nu^3)}{d\lambda} =\frac{j_\nu}{\nu^2} -(\nu\alpha_\nu)\frac{I_\nu}{\nu^3},
\end{eqnarray}
where $I_\nu$ is the intensity of the emission at frequency $\nu$, $j_{\nu}$ and $\alpha_{\nu}$ are the corresponding emissivity and absorbtivity coefficients, and $\lambda$ is the affine parameter along the photon trajectory. The synchrotron emissivity for ultra-relativistic electrons is discussed in \cite{Rybicki} deriving an exact result.  A useful  analytical approximation was obtained in \cite{Leung:2011} taking the form

\begin{align}\label{synch}
		j_\nu^\mathrm{approx}=n_e\frac{\sqrt{2}\pi e^2\nu_s}{3cK_2\left(\Theta_\mathrm{e}^{-1}\right)}\left(X^{1/2}+2^{11/12}X^{1/6}\right)^2e^{-X^{1/3}}.
\end{align}
In this expression we denote by $\Theta_e=k_BT_e/mc^2$ the dimensionless electron temperature, $K_2$ is the modified Bessel function of the second kind, and the quantities $\nu_s$ and $X$ are defined by means of the electron cyclotron frequency $\nu_{\text{cyclo}} =eB/2\pi m c$ as

\begin{align}
		X=\frac{\nu}{\nu_s},\qquad
	\nu_s=\frac{2}{9}\nu_\mathrm{cyclo}\Theta_e^2\sin{\theta}.
\end{align}.

In our work we use this approximation and further average the emissivity over the emission direction $\theta$

\begin{align}
	\langle j_\nu^\mathrm{approx}\rangle=\frac{1}{4\pi}\int j_\nu^\mathrm{approx}d\Omega
	=\frac{1}{2}\int_0^\pi j_\nu^\mathrm{approx}\sin{\theta} d\theta.
\end{align}

For the typical values of the electron temperature, emission frequency and magnetization used in this paper to model the radiation from the inner region of the accretion disk around M87$^*$ the approximation given by eq. ($\ref{synch}$) provides a good estimate and  deviates from the exact values by only a few percent \cite{Leung:2011}.

The absorption coefficient for the synchrotron radiation follows from the emissivity by the Kirchhoff's law

\begin{align}
	\alpha_\nu=\frac{j_\nu}{B_\nu(T_e)},
\end{align}
where $B_{\nu}(T_e)$ is the Planck function at the electron temperature $T_e$. For the frequencies which we consider the Rayleigh-Jeans approximation is valid leading to
\begin{align}
	\alpha_\nu\approx\frac{c^2}{2k_BT_e\nu^2}j_\nu.
\end{align}

We further need to specify a model for the dynamics of the emitting electrons in the accretion disk. We assume purely orbital motion following the prescription given in \cite{Gold:2020},\cite{Vincent:2022}. The four-velocity of the emitting electrons possesses the form

\begin{align}
        u_\mu d x^\mu&=u_0\left(-d t+\ell d\phi\right),\qquad
    \ell=\frac{\rho^{3/2}}{1+\rho},
\end{align}
where the polar and radial components vanish. In order to normalize the four-velocity so that $u_\mu u^{\mu}=-1$ is satisfied we require that
\begin{align}
	u_0=\frac{1}{\sqrt{-\left(g^{tt}-2g^{t\phi}\ell+g^{\phi\phi}\ell^2\right)}}.
\end{align}

\section{Accretion disk images}

In this section we construct the images of the accretion disks around the Kerr black hole and the two types of naked singularities using the phenomenological RIAF accretion model which we previously described. First we present simulations of the accretion disk images as they theoretically appear to a distant observer. However, due to the limited resolution of the EHT and the sparse covering of the telescope arrays the realistic images differ substantially from the theoretical predictions. Therefore, in the following subsections we reconstruct the observable images through the current EHT array exploring their most essential features at the corresponding resolution. The analysis is based on the 2017 EHT array and the impact of near-future developments as extending the telescope arrays or observing at a second frequency are considered in the next section.

\subsection{Simulated images}

The observable image of the accretion disk can be obtained theoretically by performing a ray-tracing procedure. We integrate numerically the null geodesic equations for trajectories originating from the disk as described by eq. ($\ref{disk_model}$) and reaching a distant observer. We assume that the disk emission is produced by a synchrotron radiation with intensity determined by eq. ($\ref{synch}$). The model is adapted to the characteristics of the galactic target M87$^*$. We assume an inclination angle $\theta = 160^\circ$. The mass of the compact object is $M=6.2\times 10^9 \, \text{M}_{\odot}$  while its distance is$D= 16.9$ Mpc taking into account estimates considered in \cite{EHT5}.

We perform the ray-tracing simulations for the Gauss-Bonnet and the Janis-Newman-Winicour naked singularities taking the representative values of the scalar field parameter $\gamma = 1.15 $ and $\gamma = 0.48$, respectively. As a reference we provide the corresponding disk images for the Kerr black hole. The effect of the black hole spin on the image features is explored by considering non-spinning black holes, as well  rotating black holes with a spin parameter $a=0.5$.

\begin{figure}[t!]
\centering
  \includegraphics[width=0.8\textwidth]{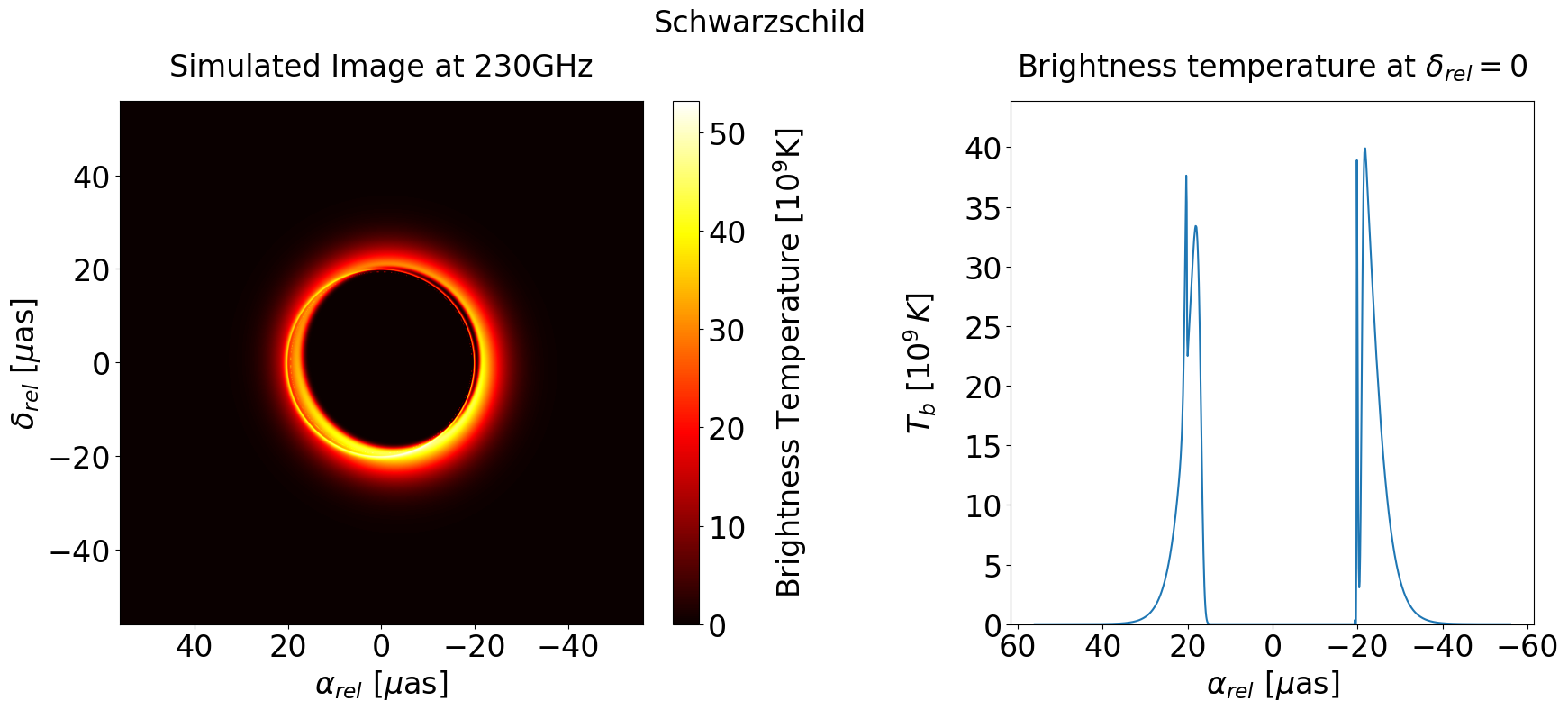}\\
  \includegraphics[width=0.8\textwidth]{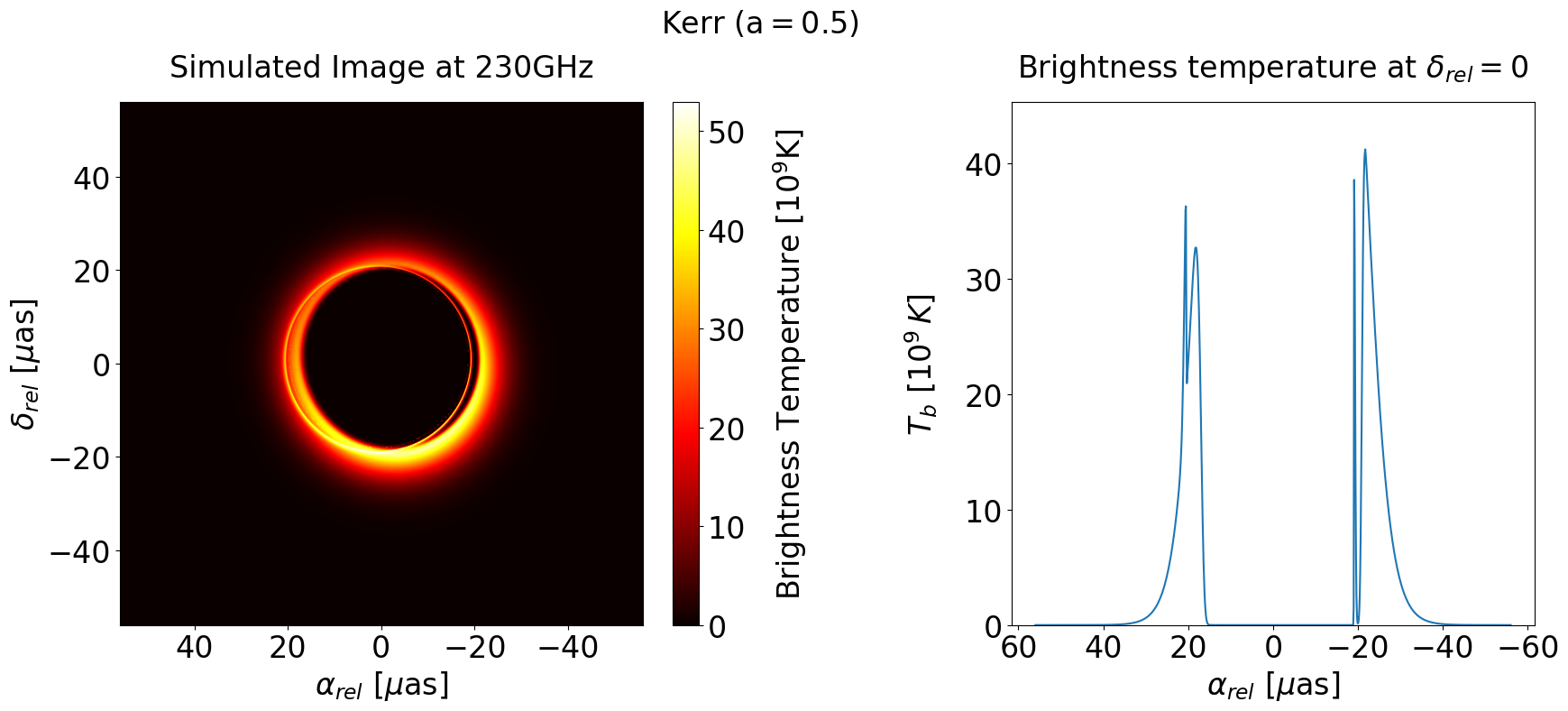}
 \caption{Simulated images for the Schwarzschild black hole and the Kerr black hole with the spin parameter $a=0.5$. The cut-off radius is $r_0 = 4.5M$, while the inner temperature is $T_0=6.5\times 10^{10}$ K for both solutions. The total flux of the image is $F=0.574$ Jy for the Schwarzschild black hole and $F=0.582$ Jy for the Kerr solution. For the remaining parameters of the accretion disk model and the ray tracing procedure see Table $\ref{table:Ray_tr}$.}
\label{fig:Ray_tr_Kerr}
\end{figure}

\begin{figure}[t!]
\centering
   \includegraphics[width=0.8\textwidth]{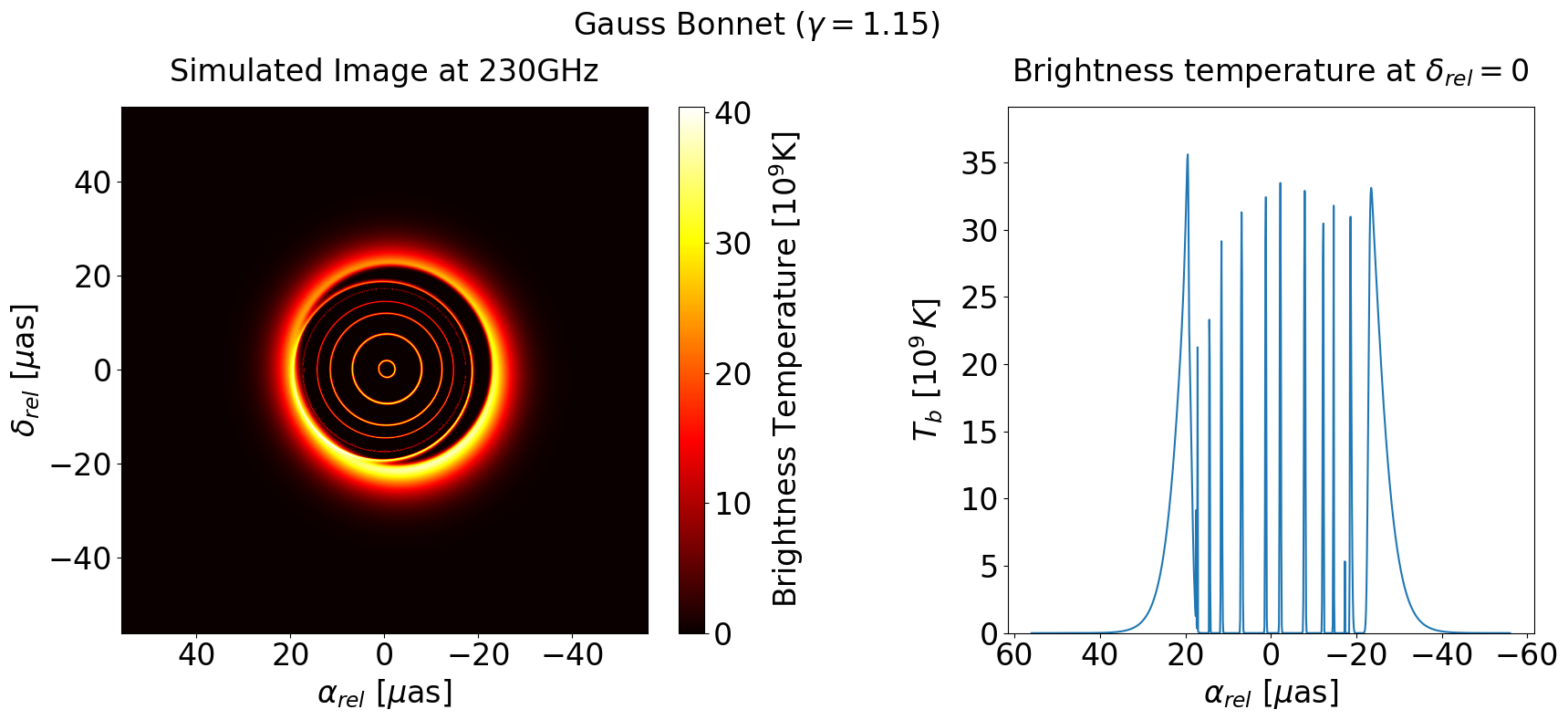}\\
  \includegraphics[width=0.8\textwidth]{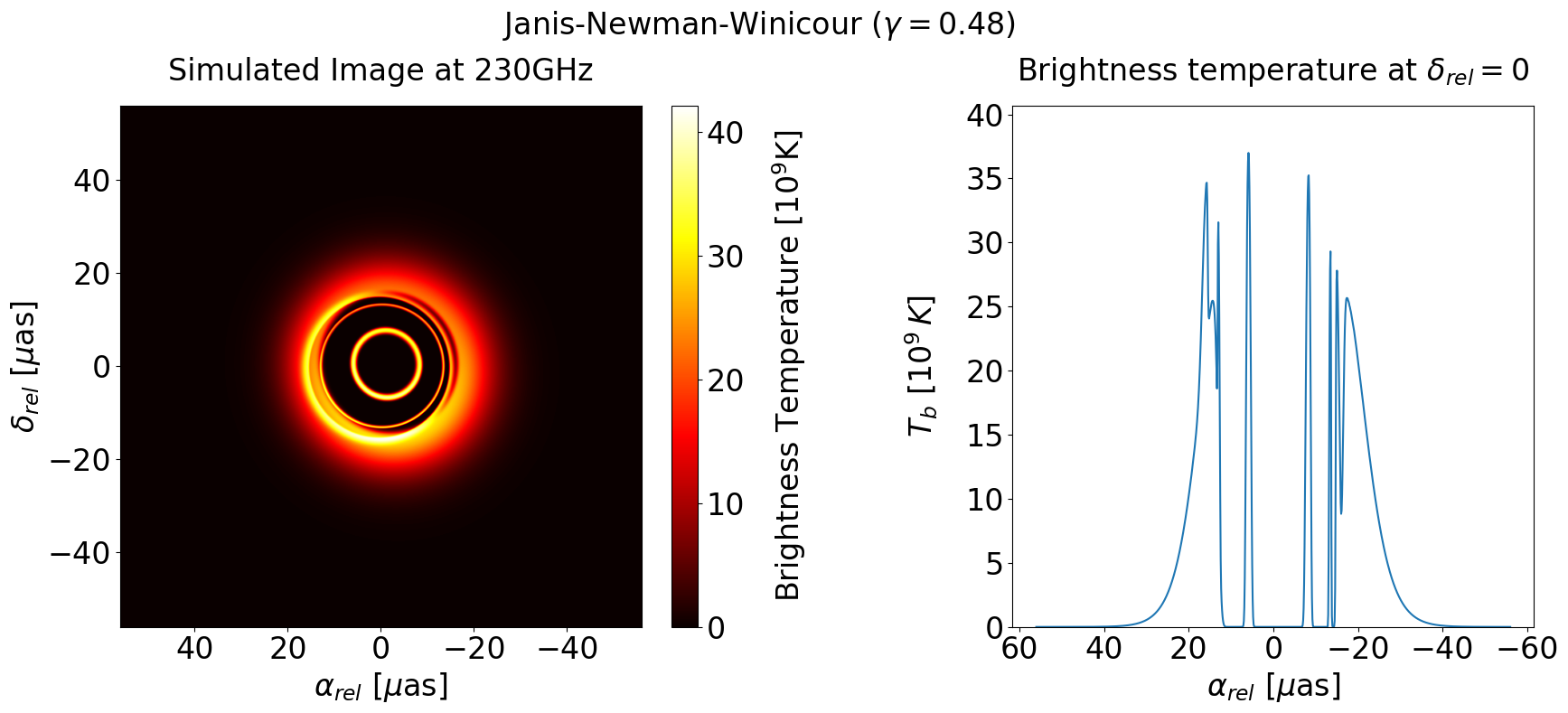}
 \caption{Simulated images for the Gauss-Bonnet naked singularity with a coupling constant $\gamma=1.15$ and the JNW naked singularity with scalar field parameter $\gamma=0.48$. The cut-off radius is $r_0 = 5M$ for both solutions, while the inner temperature is $T_0=5.9\times 10^{10}$ K for the Gauss-Bonnet solution and $T_0 = 7.2\times 10^{10}$ K for the JNW spacetime. The total flux of the image is $F=0.545$ Jy for the Gauss-Bonnet naked singularity and $F=0.597$ Jy for the JNW solution. For the remaining parameters of the accretion disk model and the ray tracing procedure see Table $\ref{table:Ray_tr}$.}
\label{fig:Ray_tr_NS}
\end{figure}

The results are presented in Figs. $\ref{fig:Ray_tr_Kerr}$-$\ref{fig:Ray_tr_NS}$ where we also plot the variation of the brightness temperature along the horizontal cross-section of the disk image at $\delta_{rel}=0$. The disk model parameters are summarized in Table $\ref{table:Ray_tr}$ where the cut-off radius $r_0$  and the inner temperature $T_0$ vary for the different compact objects since they are fine-tuned in each case to produce an angular size of the disk image and total flux compatible with the measurements for M87$^*$.

\begin{table}[h!]
\centering
\begin{tabular}{c|c}
            \hline
		{parameter} & {value}
		\\\hline\hline
		compact object mass (M) & $6.2 \times 10^{9} M_{\odot}$
		\\
		compact object distance (D) & 16.9 Mpc
		\\
		disk opening angle ($\alpha=\tan\theta_0$)  & 0.1
		\\
		electron number density  at the cut-off radius ($n_0$) & $5\times 10^{5}$  cm$^{-3}$
		\\
		magnetization ($\sigma$) & 0.01
		\\
		disk sharpness parameter ($r_{sc}$)  & 0.4M
		\\
		inclination angle  ($i$)  & $160^\circ$
		\\
		observational frequency ($\nu$)  & 230 GHz
		            \\ \hline
	\end{tabular}
 \caption{Parameters of the accretion disk model and the ray-tracing set-up. }
 \label{table:Ray_tr}
\end{table}

We see that the naked singularities lead to qualitatively different morphology of the disk images. They produce a series of bright rings at the center of the image which are absent for the Kerr black hole where we observe a pronounced depression of the image brightness. For the Janis-Newman-Winicour naked singularity the central  bright rings are formed due to the capacity of the spacetime to produce double images of order up to $k=2$ for axisymmetric matter distributions \cite{Nedkova:2020}. In the case of the Gauss-Bonnet naked singularity the central rings occur due to the presence of an infinite sequence of double images of order $k\geq 1$ \cite{Nedkova:2021}.

The central rings radiate with significant intensity.  The brightness temperature of the innermost rings for the Gauss-Bonnet naked singularity is comparable to the maximal emission from the main disk image. For the Janis-Newman-Winicour naked singularity their radiation exceeds the radiation from the main disk representing the brightest feature in the image. The morphology of the central ring could  be difficult to be observed with the current resolution of the EHT. However,  the emission from the central part of the image is considerable. Therefore,  we expect that it will be encoded in a certain quantitative feature which can serve as a measure for distinguishing naked singularities from black holes. In the next subsections we consider the observational impact of the central bright rings on the disk images using the 2017 EHT array and define some image metrics which can quantify it.

\subsection{Reconstructed images}

In order to able to estimate the observational features of the naked singularities spacetimes in comparison to the Kerr spacetime we should simulate the accretion disk images as seen by the Event Horizon Telescope array. For the purpose we use a  maximum likelihood reconstruction method for interferometric imaging which was developed in \cite{Chael:2018} and implemented in the open access software package ehtim\footnote{https://github.com/achael/eht-imaging}. This approach obtains the source image by means of performing a best-fit procedure to certain data products constrained by additional regularization terms. In particular the method minimizes the objective function $J(I)$

\begin{eqnarray}
    J(I)= \sum_{\text {data terms}}\alpha_D \chi_D^2(I) -\sum_{\text {regularizers}} \beta_R S_R(I),
\end{eqnarray}
where $\chi^2_D$ are the goodness-of-fit functions associated with the data quantity $D$, and $S_R$ are regularizing functions. The $\chi^2_D$ terms characterize the probability that the corresponding data is observed given a particular image $I$. The regularizing functions $S_R$ constrain the set of images which could be associated with the measured data by providing additional information about the image properties. We consider four  regularizing functions which restrict the entropy and the smoothness of the images  as well as provide constraints on the total image flux density and its position in the field of view.

The maximum likelihood method allows to use data quantities formed as products of the complex visibilities which reduce the station-based errors in the measured data. In this work we use the closure phases which are defined as the phases of the bispectra formed by multiplying any three visibilities located at a triangle. The closure phases eliminate any station-based phase errors resulting in measured quantities which coincides with the observed values. As a second data term we consider the visibility amplitudes following \cite{EHT4}.

The contribution of the different data and regularizer terms in the objective function $J(I)$ is controlled by the weight factors $\alpha_D$ and $\beta_R$. Their values are chosen accordingly to facilitate convergence of the minimization procedure and lead to acceptable values of the goodness-of-fit parameters $\chi^2_D$. One of the parameters can be interpreted as a scale factor and set to unity since the global minimum of $J(I)$ does not depend on scale. In our simulations we use the parameter values reported in \cite{Chael:2018}. Each image is obtained by performing a four-stage iteration procedure as the final image from each round is blurred and used as an initial image for the following round. The maximum number of iterations in each round is set to $N_{\text{iter}}= \{1000, 3000, 4000, 4000\}$, respectively. Since the image reconstruction procedure leads to resolution which exceeds the realistic one, the final image is blurred with a Gaussian which is proportional to the interferometer clean beam but scaled by a factor $1/2$.

\begin{figure}[t!]
\centering
  \includegraphics[width=0.9\textwidth]{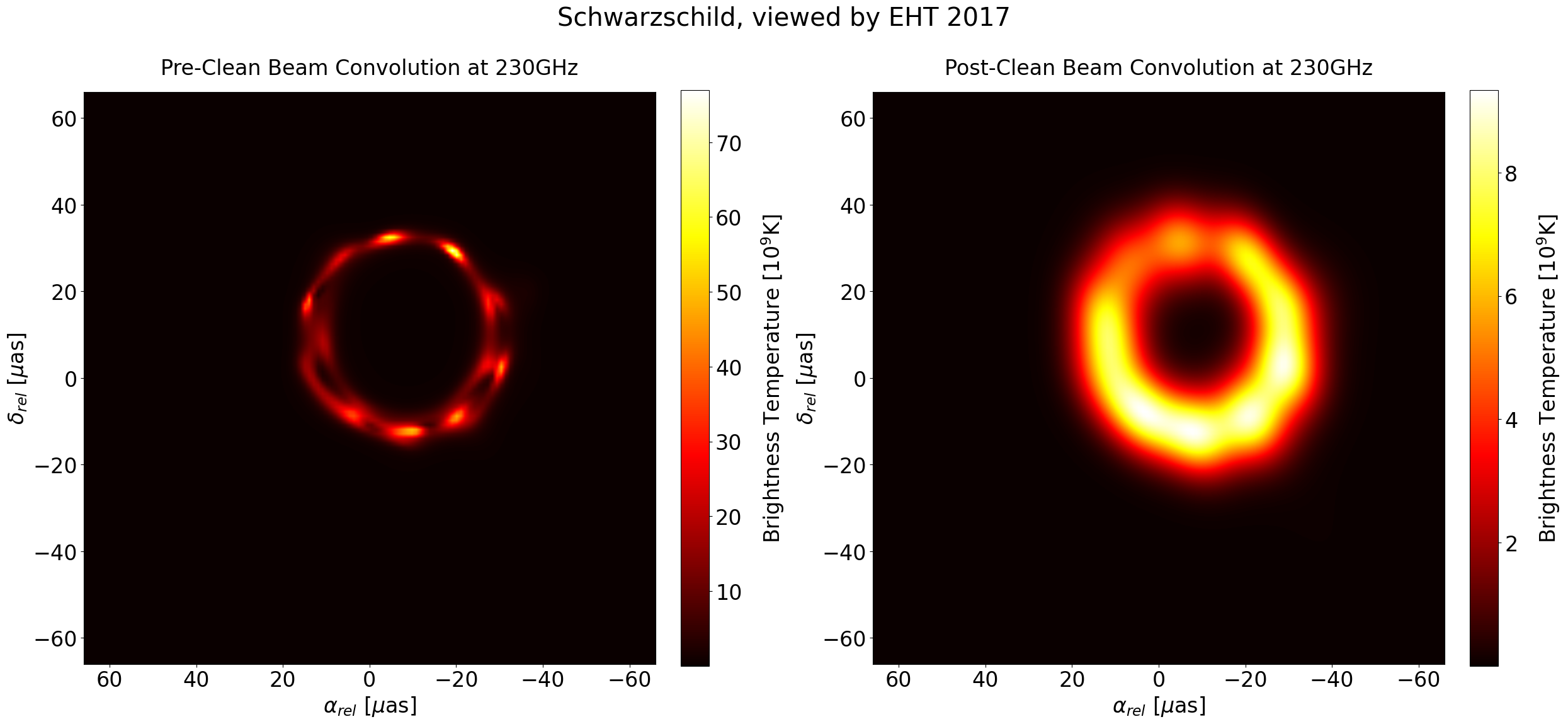}\\
  \includegraphics[width=0.9\textwidth]{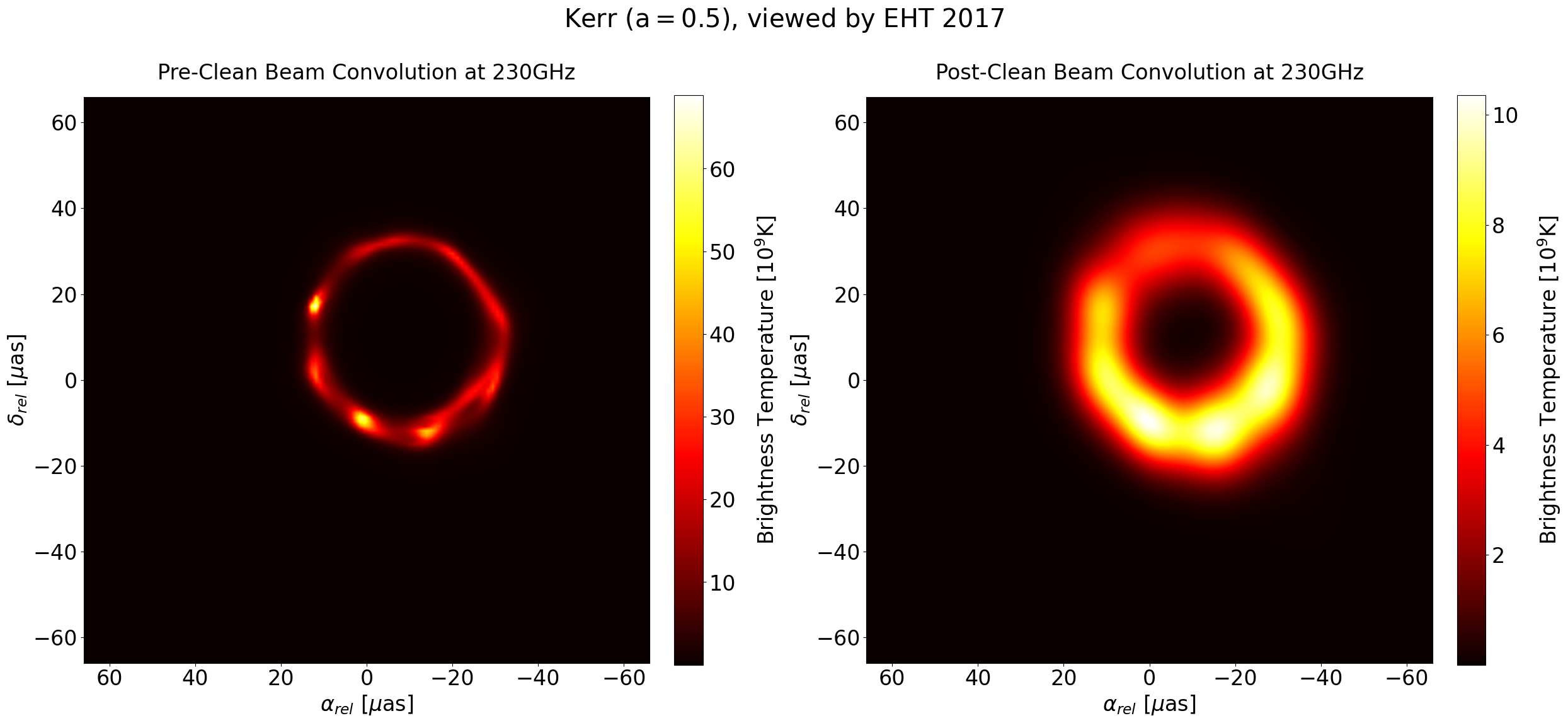}
 \caption{Reconstructed images for the Schwarzschild black hole and the Kerr black hole with the spin parameter $a=0.5$. The left panel provides the image at full resolution while the image on the right is blurred with $1/2$ interferometer clean beam. The goodness-of-fit parameters associated with the visibility amplitudes and the closure phases are $\chi^2_{\text{vis}}=1.02$ and $\chi^2_{\text{cl}}=0.90$ for the Schwarzschild black hole and $\chi^2_{\text{vis}}=1.04$ and $\chi^2_{\text{cl}}=1.07$ for the Kerr black hole.}
\label{fig:ehtim_Kerr}
\end{figure}

\begin{figure}[t!]
\centering
   \includegraphics[width=0.9\textwidth]{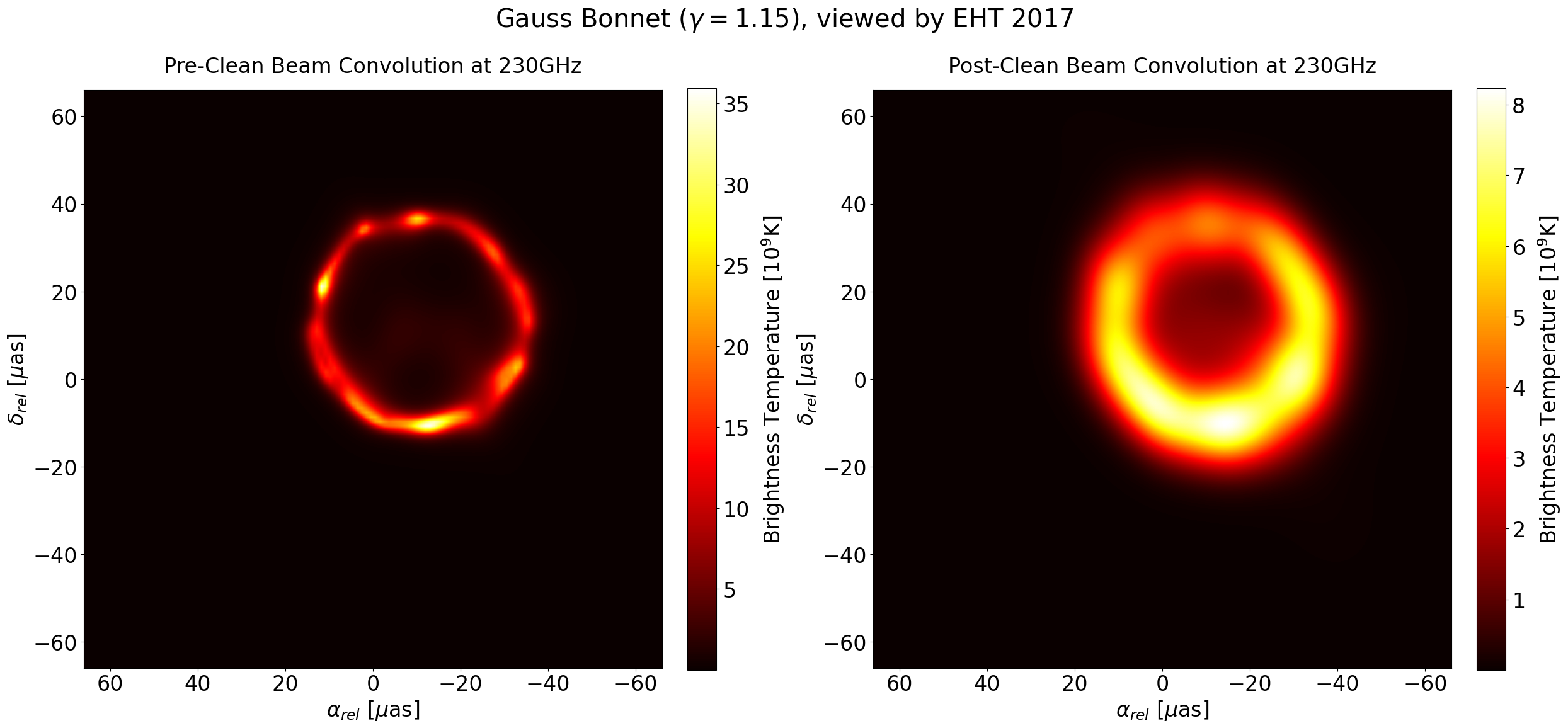}\\
  \includegraphics[width=0.9\textwidth]{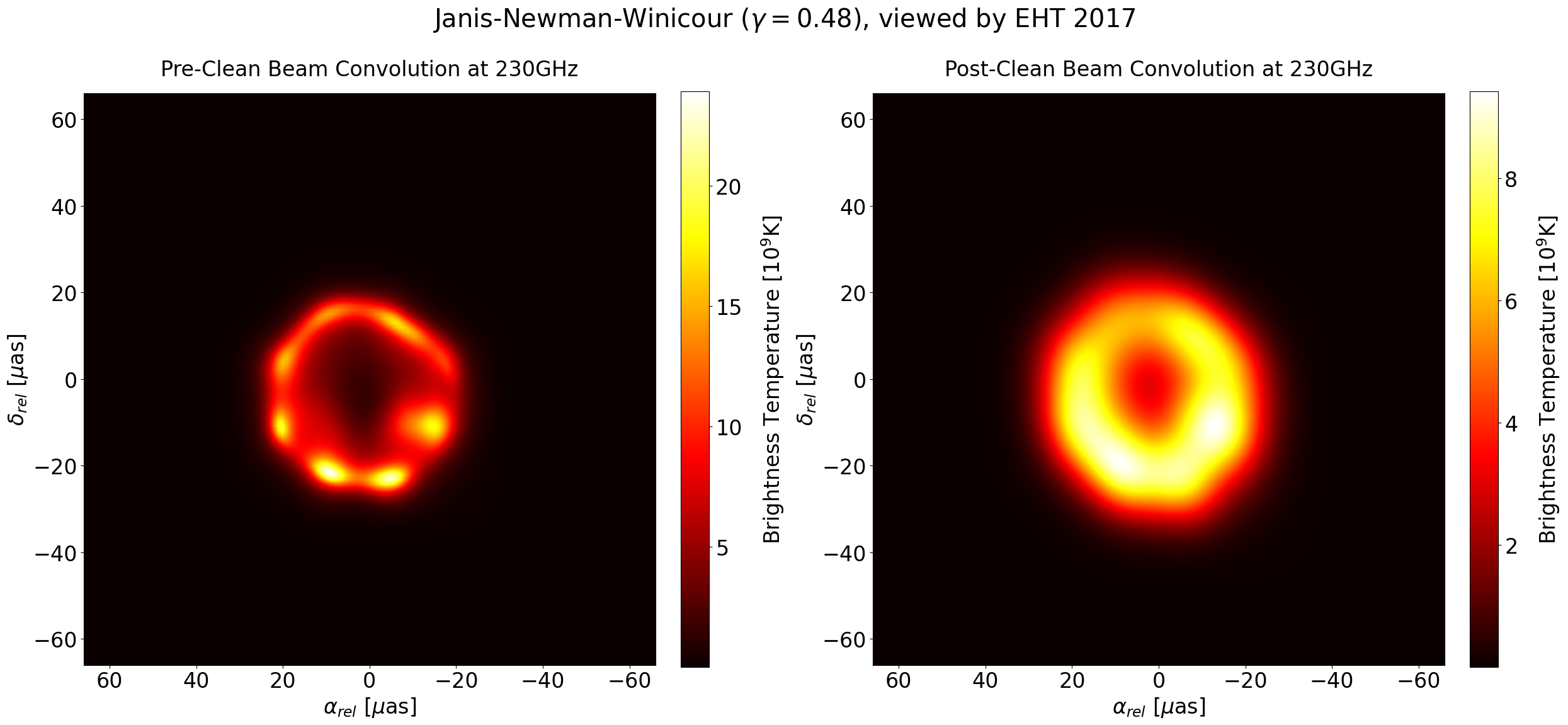}
 \caption{Reconstructed images for the Gauss-Bonnet naked singularity with a coupling constant $\gamma=1.15$ and the JNW naked singularity with scalar field parameter $\gamma=0.48$. The left panel provides the image at full resolution while the image on the right is blurred with $1/2$ interferometer clean beam. The goodness-of-fit parameters associated with the visibility amplitudes and the closure phases are $\chi^2_{\text{vis}}=1.01$ and $\chi^2_{\text{cl}}=0.91$ for the Gauss-Bonnet naked singularity, and  $\chi^2_{\text{vis}}=1.00$ and $\chi^2_{\text{cl}}=0.84$ for the JNW solution.}
\label{fig:ehtim_NS}
\end{figure}

In Figs. $\ref{fig:ehtim_Kerr}$-$\ref{fig:ehtim_NS}$ we present the reconstructed images corresponding to the simulated data obtained in Figs. $\ref{fig:Ray_tr_Kerr}$-$\ref{fig:Ray_tr_NS}$. We consider the Event Horizon Telescope array which was used in the observations in 2017 giving the detailed configuration in Appendix A.  In each figure we provide the values of the goodness-of-fit parameters $\chi^2_{vis}$ and $\chi^2_{cl}$ associated with the visibility amplitude and the closure phase data terms, respectively.

We see that due to the limited resolution the internal structure at the center of the naked singularity images is blurred and the disk morphology looks qualitatively similar to the Kerr black hole. However, we can further notice that the interior of the ring image is significantly brighter for the naked singularities than for the Kerr black hole. We can quantify the emission from the central depression and define an appropriate measure which can distinguish naked singularities from black holes. We perform this analysis in the next subsection.

\subsection{Template-based image properties }

In order to be able to describe accretion disk images quantitatively we should introduce certain measures evaluating their geometry and brightness. In the EHT Collaboration analysis of the M87* data this issue is approached by introducing a ring template and performing a best-fit procedure of the disk image to it \cite{EHT4}. The ring template is characterized by its diameter, width   and orientation and the best-fit values of these parameters are assigned as geometrical characteristics of the observed image. In our work we will use a similar procedure for extracting the image features which is implemented in the open source software package VIDA \cite{Tiede:2022}\footnote{https://github.com/ptiede/VIDA.jl}.

VIDA defines a ring template which uses eight parameters to constrain the image geometry. The image size is parameterized by the ring diameter $d_0$ and the region of substantial radiation is associated with the width of the ring $w$. The ring is not supposed to be circular since asymmetries may be introduced in the reconstruction of the EHT image. Thus, we consider an elliptic distribution which is centered at $(x_0, y_0)$ and possesses semi-major and semi-minor axes $a$ and $b$, respectively. Then, we can define the simplest template as a Gaussian ring with diameter $d_0=2\sqrt{ab}$ and  width $w = 2\sqrt{2\ln2}\sigma$ determined by the FWHM of the Gaussian with standard deviation $\sigma$.  We further denote the ellipticity of the ellipse by $\tau=1-b/a$ and the position angle of the semi-major axis by $\xi_{\tau}$ which is measured north of east. Finally, we should take into account the asymmetry of the emission around the ring caused by the Doppler boost. For the purpose a slash function is introduced depending on the azimuthal angle $\phi$ around the ring

\begin{equation}\label{eq:slash}
    S(x,y; s,\xi_s) = N_0\left[1+s\cos(\phi-\xi_s)\right].
\end{equation}
It is parameterized by the strength of the slash $s$ and its position angle $\xi_s$ measured east of north, while $N_0$ is a normalizing factor ensuring unit normalization. In this way we obtain the following template

\begin{equation}\label{ring}
    h_{\theta}(x,y) = S(x,y; s,\xi)\exp\left[ - \frac{(d_\theta(x, y))^2}{2 \sigma^2} \right],
\end{equation}
where $d_\theta(x, y)$ is the minimum distance between the ellipse with parameters $\theta = \{d_0, \tau, x_0, y_0\}$ and the point $(x, y)$.

The image features are extracted by performing a minimization procedure using the Bhattacharyya divergence as an objective function. The Bhattacharyya divergence is defined as

\begin{equation}
    Bh(I(x,y)||h(x,y)) = -\log\int\sqrt{I(x,y)h(x,y)}dxdy,
\end{equation}
comparing the template distribution $h(x,y)$ to the image intensity $I(x,y)$. The template parameters which minimize the Bhattacharyya divergence describe most closely the image geometry.

\begin{figure}[t!]
\centering
  \includegraphics[width=\textwidth]{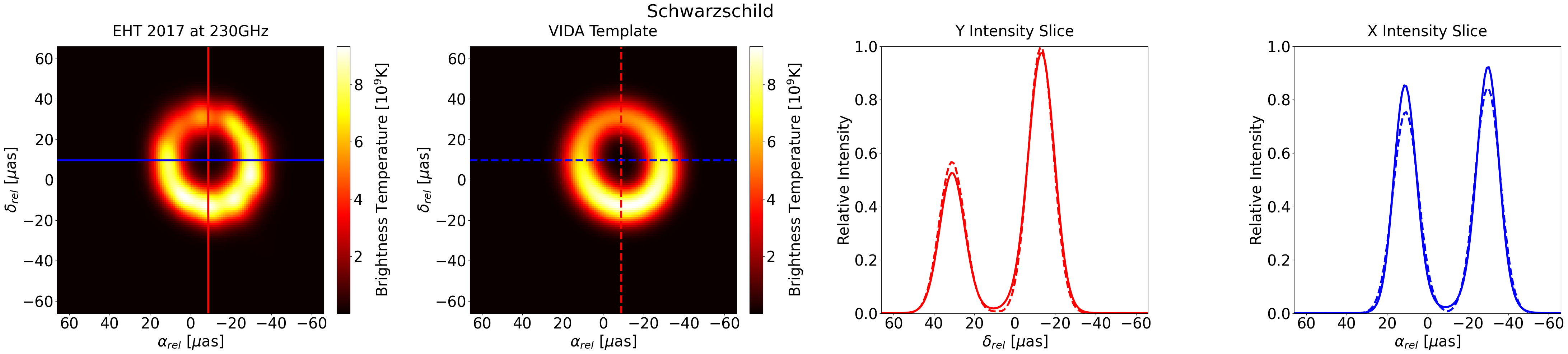}\\[3mm]
  \includegraphics[width=\textwidth]{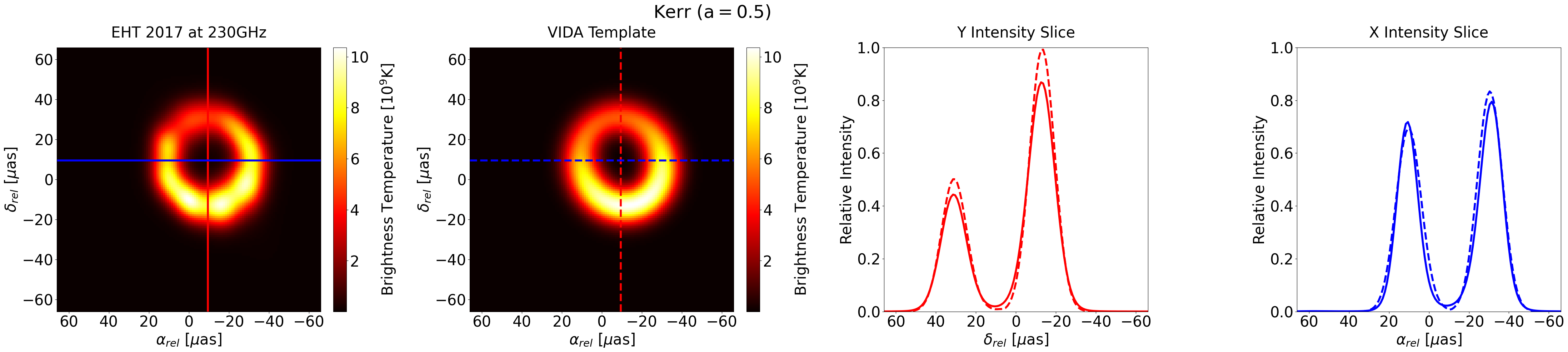}\\[3mm]
  \includegraphics[width=\textwidth]{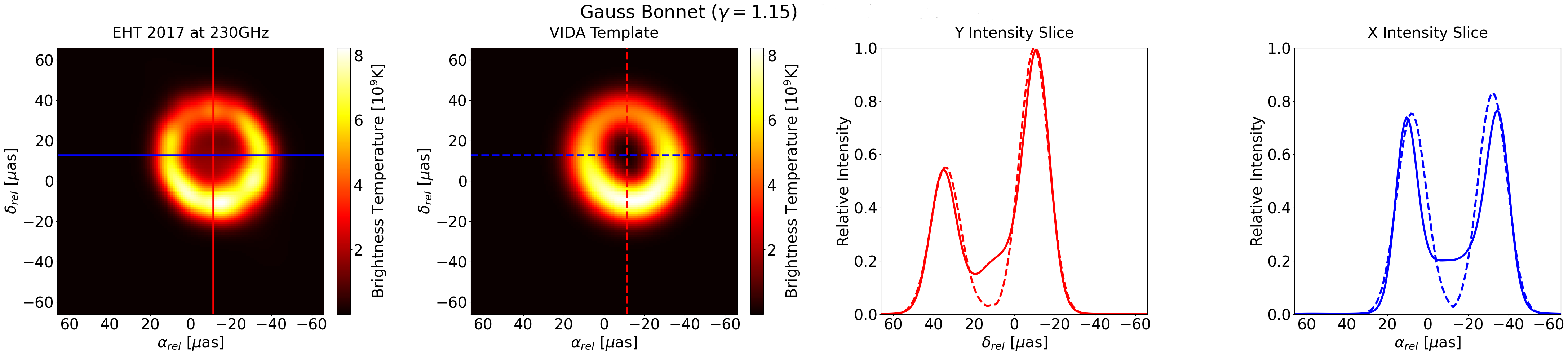}\\[3mm]
   \includegraphics[width=\textwidth]{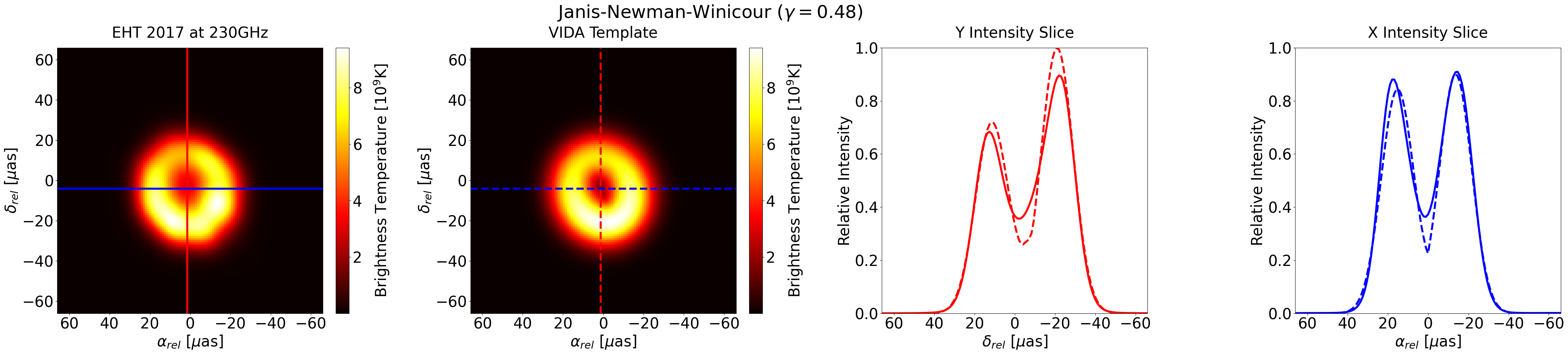}
 \caption{Reconstructed images (left) and VIDA fit with a ring template (right) for the accretion disks in the Gauss-Bonnet and Janis-Newman-Winicour naked singularities spacetimes. The corresponding images for the Schwarzschild and Kerr black holes are presented as a reference. In the right panel we investigate the  variation of the radiation intensity across the horizontal  (blue line) and  vertical (red line) cross-section through the center of the fitted ellipse. The intensity cross-sections are compared for the reconstructed images (solid lines) and the fitted VIDA templates (dashed lines). }
\label{fig:VIDA_2017}
\end{figure}

We perform the described procedure on the reconstructed images in the previous section \footnote{ We apply the VIDA toolkit to perform the variational analysis and extract the image features. However, as a final step we use our own module to visualize the results.}. Our results are presented in Fig. $\ref{fig:VIDA_2017}$ and we summarize the best-fit parameters and the optimized divergence in Table $\ref{table:VIDA_2017}$. In order to evaluate how well the fitted templates reproduce the radiation intensity of the reconstructed images we compare the variation of the intensity across the horizontal and the vertical cross-sections through the center of the fitted ellipse $(x_0, y_0)$.

We see that for the Einstein-Gauss-Bonnet naked singularity the image characteristics are very similar to those of the Kerr black hole. Furthermore, they are consistent with the best-fit parameters reported in the analysis of the M87$^*$ image in \cite{EHT4} within the evaluated limits. Thus, the Einstein-Gauss-Bonnet naked singularity may lead to observable images of the accretion disk with a ring diameter and width compatible with the geometry of the M87$^*$ image in the same way as the Kerr black hole.

The Janis-Newman-Winicour singularity leads to a smaller diameter of the accretion disk image than the reported values in the analysis of the M87* data. The image diameter depends substantially on the properties of the  accretion model. In our case it is most strongly influenced by the value of the cut-off radius $r_0$. However, it is also constrained by the spacetime geometry by means of its lensing properties. In spacetimes possessing a photon sphere it is related to the size of the shadow, i.e. the lensed image of the photon ring. It was demonstrated in \cite{Kocherlakota:2021} that weakly naked Janis-Newman-Winicour singularities with scalar field parameter $\gamma<0.53$ lead to diameters of the observable ring image which are smaller than the compatibility interval for the M87$^*$ data. This can be explained by the stronger focusing effect of the Janis-Newman-Winicour spacetime which becomes more pronounced when the scalar field parameter decreases \cite{Virbhadra:2002}, \cite{Nedkova:2019}. The same argument can be applied to the naked singularities with $\gamma=0.48$ which we consider in this work.  By fine-tuning the parameters of the accretion model we were unable to obtain a ring diameter  which fits into the compatibility interval with the M87$^*$ data. Therefore, we conclude that the smaller disk image size in this case is an intrinsic feature of the underlying spacetime which is determined mostly by its lensing properties rather than the specifics of the accretion model.

\begin{table}
\centering
\begin{tabular}{c|c|c|c|c}
            \hline
		{template parameter} & {Schw.}&{\thead{Kerr \\ (a=0.5)}}&{Gauss-Bonnet}&{JNW}
		\\\hline\hline
		$\sigma$ {(ring width [$\mu$as])} & 6.17&6.21&7.26&7.94
		\\
		$\tau$ {(ellipticity)} & 0.10&0.10&0.18&0.15
		\\
		$\xi_\tau$ {(ellipse orientation)} & -1.95&1.19&-1.96&-1.95
		\\
		$s$ {(slash)} & 0.28&0.34&0.29&0.17
		\\
		$\xi_s$ {(slash orientation)} & -1.78&-1.85&-1.73&-1.77
		\\\hline
		$r_0$ {(ring radius [$\mu$as])} & 21.1&21.1&21.0&15.4
		\\
		$x_0$ {(offset RA [$\mu$as])} & -8.78&-9.47&-11.37&1.38
		\\
		$y_0$ {(offset DEC [$\mu$as])} & 9.61& 9.46&12.68&-4.10
		\\\hline\hline
		{  optimized divergence} & 0.005&0.006&0.007&0.004
            \\ \hline
	\end{tabular}
 \caption{Best-fit parameters of the VIDA ring template for the accretion disk images in the Gauss-Bonnet and Janis-Newman-Winicour naked singularities spacetimes. As a comparison we perform the same analysis for the Schwarzschild and Kerr black hole with spin parameter $a=0.5$. }
 \label{table:VIDA_2017}
\end{table}

The deviation of the naked singularity disk images from the Kerr black hole can be quantified by introducing a measure for the central brightness depression \cite{EHT4}, \cite{Eichhorn:2023}. Using the ring template ($\ref{ring}$) we can separate the image into two sets defined as a ring region $\cal{R}$ and a shadow region $\cal{S}$. The ring region consists of the image points $(x, y)$ which are located within a distance $1\sigma$ from the best-fit ellipse $(\tau, \xi_\tau)$.

\begin{align}
	\label{eq:ringRegion}
	\mathcal{R}=\left\lbrace
		(x,y)\in\mathbb{R}^2\;:\;
		d_\theta(x,y;\,d_0,\tau,\xi_\tau,x_0,y_0)\leqslant\sigma
	\right\rbrace\;.
\end{align}

On the other hand, the shadow region includes all the points located in the interior of the region $\cal{R}$. Then, we can introduce the following measure for the central  brightness depression

\begin{align}
	\label{brightness_ratio}
	\hat{f}_{c} = \frac{\text{minimum flux in }\mathcal{S}}{\text{mean flux in }\mathcal{R}}\;,
\end{align}
which is evaluated using the reconstructed images. This quantity was also considered in the analysis of the properties of the M87$^*$ image in \cite{EHT4}, however using a different template obtaining the value $\hat{f}_c\, (\text{M87}^*) = 0.04$ .

We evaluate the brightness ratios for the reconstructed naked singularity and black hole images which are presented in Fig. $\ref{fig:VIDA_2017}$ summarizing the results in Table $\ref{table:f_2017}$ . We see that for the Kerr black hole we obtain similar values as the brightness ratio reported in the analysis of M87$^*$ irrespective of the spin parameter. On the other hand, for the naked singularities the brightness ratio is with an order of magnitude higher. Thus, although the fine structure in the inner region of the image is not resolved, the enhanced brightness of central depression can serve as a signature for distinguishing such types of spacetimes.

\begin{table}[t]
\centering
\begin{tabular}{||c|c|c|c|c||}
            \hline
		{spacetime} & {Schw.}&{\thead{Kerr \\ (a=0.5)}}&{Gauss-Bonnet}&{JNW}
		\\\hline
		{\thead{$\hat{f}_c$\\(brightness ratio)}} & 0.026&0.030&0.239&0.451
		\\\hline
	\end{tabular}
\caption{Brightness ratio of the central depression for the accretion disk images in the Gauss-Bonnet and Janis-Newman-Winicour naked singularities spacetimes compared to the Schwarzschild and Kerr black holes. }
\label{table:f_2017}
\end{table}

\section{Extended telescope arrays and the ngEHT}

In the next years the spacial resolution and the dynamic range of the EHT are expected to be significantly improved due to several developments. They include increasing the EHT array by adding new telescopes and introducing a second observational frequency at 345 GHz. In this section we explore how these improvements will influence the capacity of the EHT to resolve the inner structure of the naked singularities disk images and distinguish them from black holes. For the purpose we consider first the improved EHT array designed for the 2022 observation campaign which includes three more telescopes in comparison with the 2017 EHT array and observes at 230 GHz. As a next improvement we consider a tentative next-generation EHT array consisting of 21 telescopes and observing both at 230 GHz and 345 GHz. The details of the array configurations are given in  Appendix A.

\begin{table}[h!]
\centering
\begin{tabular}{||c|c|c|c|c||}
            \hline
		{spacetime} & {Schw.}&{\thead{Kerr \\ (a=0.5)}}&{Gauss-Bonnet}&{JNW}
		\\\hline
		{\thead{$\hat{f}_c$ \\ (EHT 2022)}} & 0.009&0.009&0.18&0.290
		\\\hline
        {\thead{$\hat{f}_c$ \\ (ngEHT 230 GHz)}} & 0.007&0.007& 0.21 &0.354
		\\\hline
        {\thead{$\hat{f}_c$ \\ (ngEHT 345 GHz)}} & 0.002&0.002& 0.12 &0.220
		\\\hline
        {\thead{$\hat{f}_c$ \\ (ngEHT 230 GHz $\cup$ 345 GHz)}} & 0.005&0.005& 0.20 &0.344
		\\\hline
	\end{tabular}
\caption{Brightness ratio of the central depression for the accretion disk images in the Gauss-Bonnet and Janis-Newman-Winicour naked singularities spacetimes compared to the Schwarzschild and Kerr black holes. We explore the impact of the near-future improved telescope arrays on the brightness ratio for the different spacetimes.}
\label{table:f_2022}
\end{table}

\begin{figure}[h!]
\centering
   \includegraphics[width=0.9\textwidth]{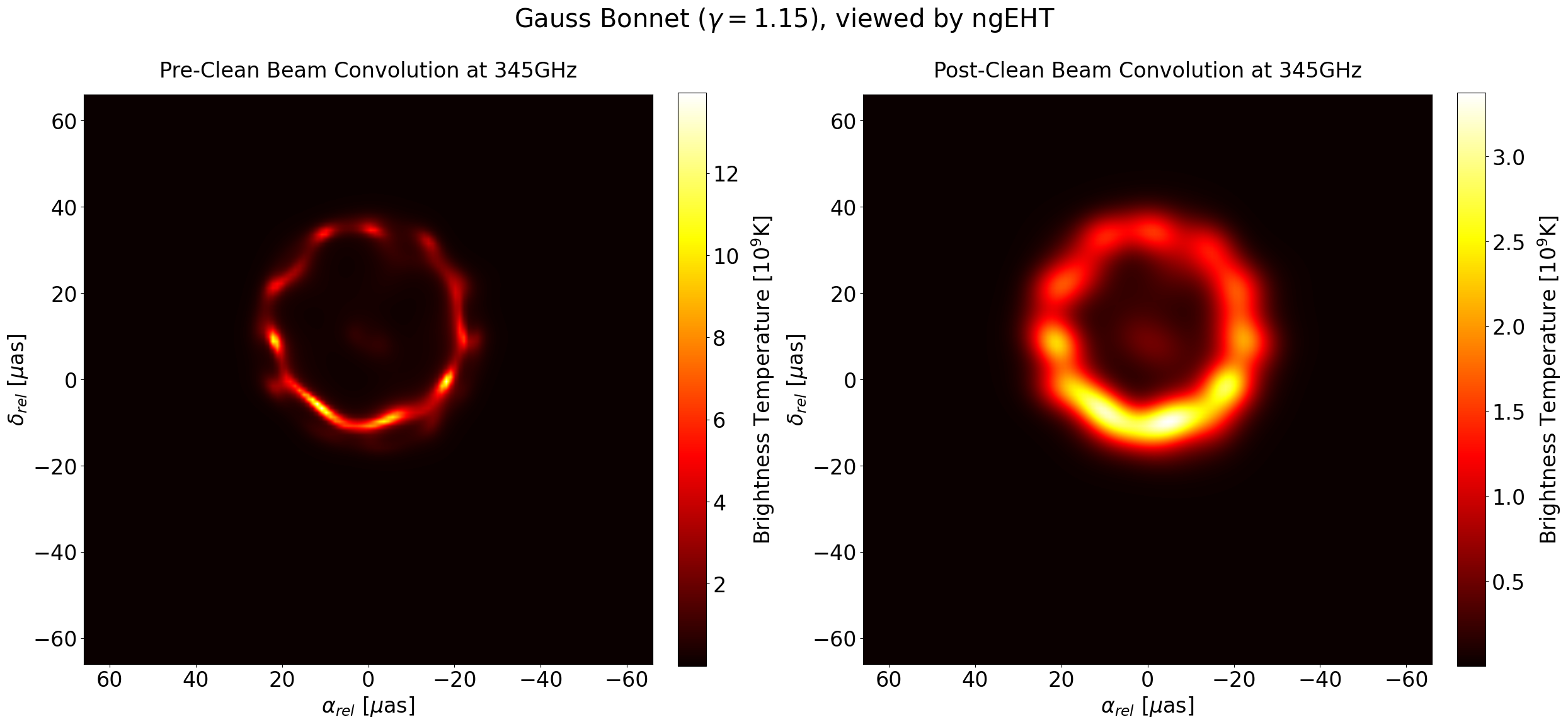}\\
  \includegraphics[width=0.9\textwidth]{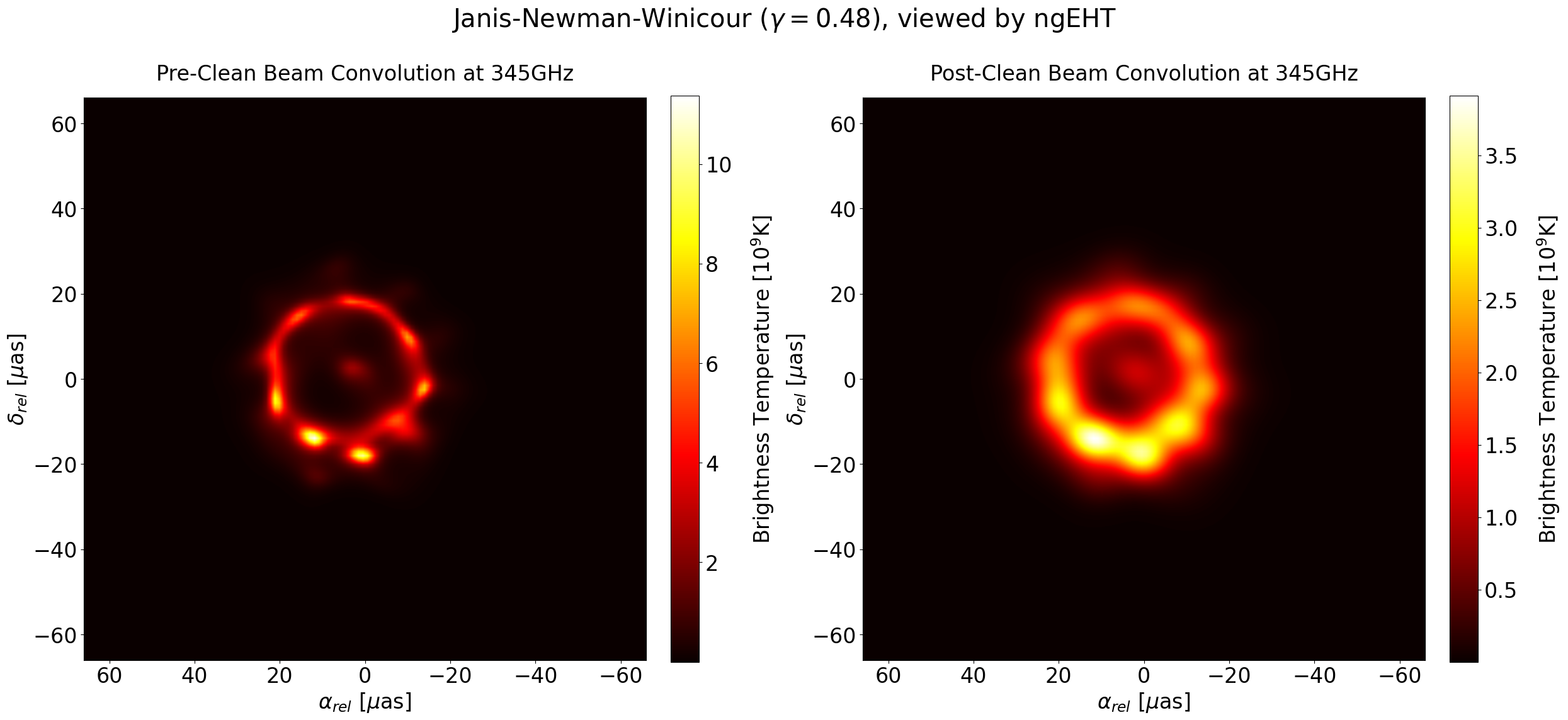}
 \caption{Reconstructed images of the accretion disks around the Gauss-Bonnet and JNW naked singularities as seen by the ngEHT array at the observing frequency $\nu=345$ GHz. The left panel provides the image at full resolution while the image on the right is blurred with $1/2$ interferometer clean beam.}
\label{fig:ehtim_ngEHT_NS}
\end{figure}

\begin{figure}
\centering
   \includegraphics[width=\textwidth]{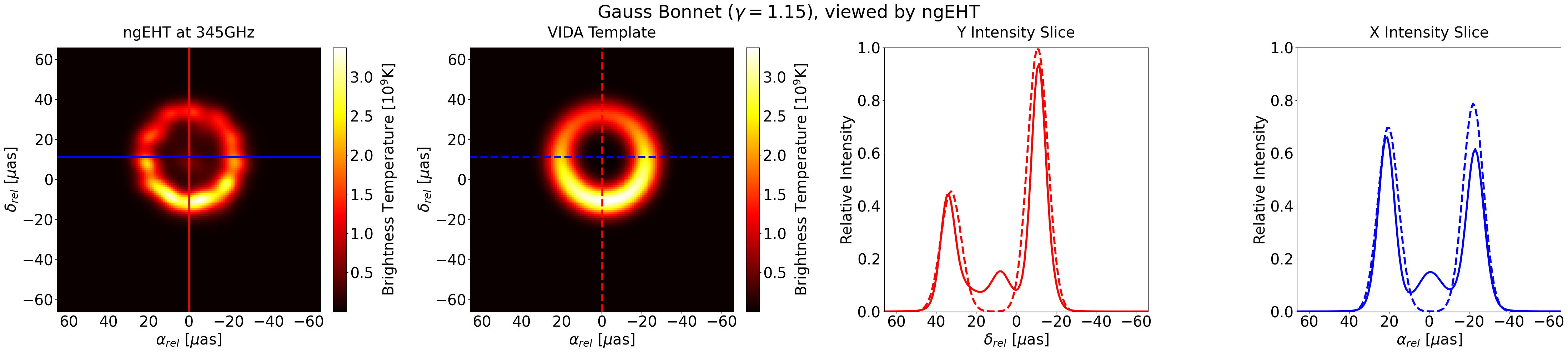}\\[3mm]
   \includegraphics[width=\textwidth]{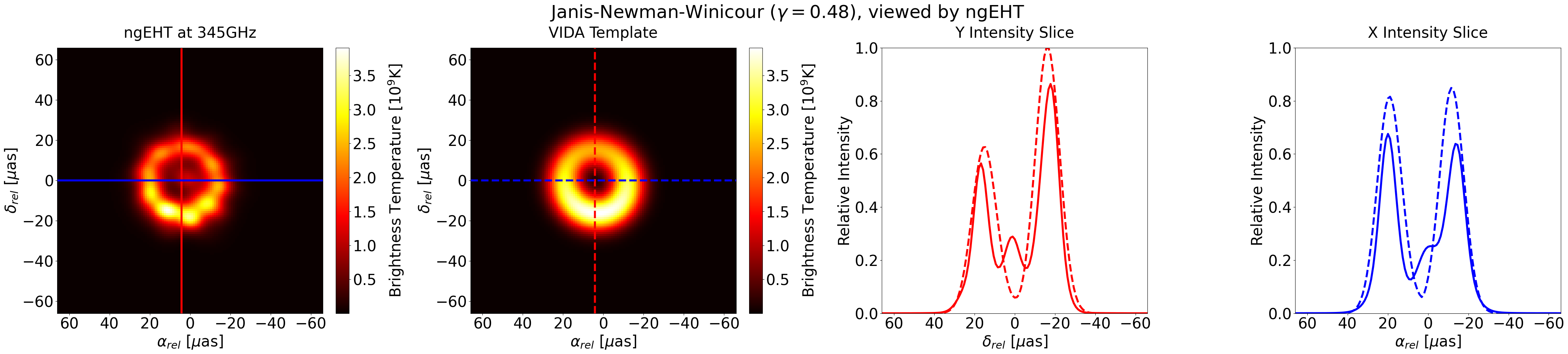}
 \caption{Reconstructed images (left) and VIDA fit with a ring template (right) for the accretion disks as seen by the ngEHT array at the observing frequency $\nu=345$ GHz. In the right panel we investigate the  variation of the radiation intensity across the horizontal  (blue line) and  vertical (red line) cross-section through the center of the fitted ellipse. The intensity cross-sections are compared for the reconstructed images (solid lines) and the fitted VIDA templates (dashed lines). }
\label{fig:VIDA_ngEHT_11}
\end{figure}

Using the new array configurations we perform the analysis described in section 4  for the simulated naked singularity and black hole images in Figs. $\ref{fig:Ray_tr_Kerr}$ and $\ref{fig:Ray_tr_NS}$. In addition,  we also simulate the theoretically observable disk intensity at the frequency $\nu=345$ GHz by performing a numerical ray-tracing. The extended ray-tracing simulations and the   reconstructed images are presented in Appendix B. We further provide in  Appendix C the details of the VIDA template analysis of the disk geometry and the best-fit parameters of the Gaussian ring template. As a result we obtain  the brightness ratio $\hat{f}_c$ defined in Eq. ($\ref{brightness_ratio}$) which characterizes the relative intensity of the central depression and summarize the value for the different spacetimes and array configurations in Table $\ref{table:f_2022}$.

We see that the improved telescope arrays reduce the brightness ratio compared to the 2017 EHT array both for the Kerr black hole and the naked singularities. This effect is expected as a result of  the increased resolution. However, in the case of black holes the ratio is decreased by an order of magnitude while for naked singularity it reduces only slightly. In this way the near-future telescope arrays will observe much greater deviations in the intensity of the central brightness depression for naked singularities spacetimes in comparison to black holes.

\begin{figure}[t!]
\centering
   \includegraphics[width=1.\textwidth]{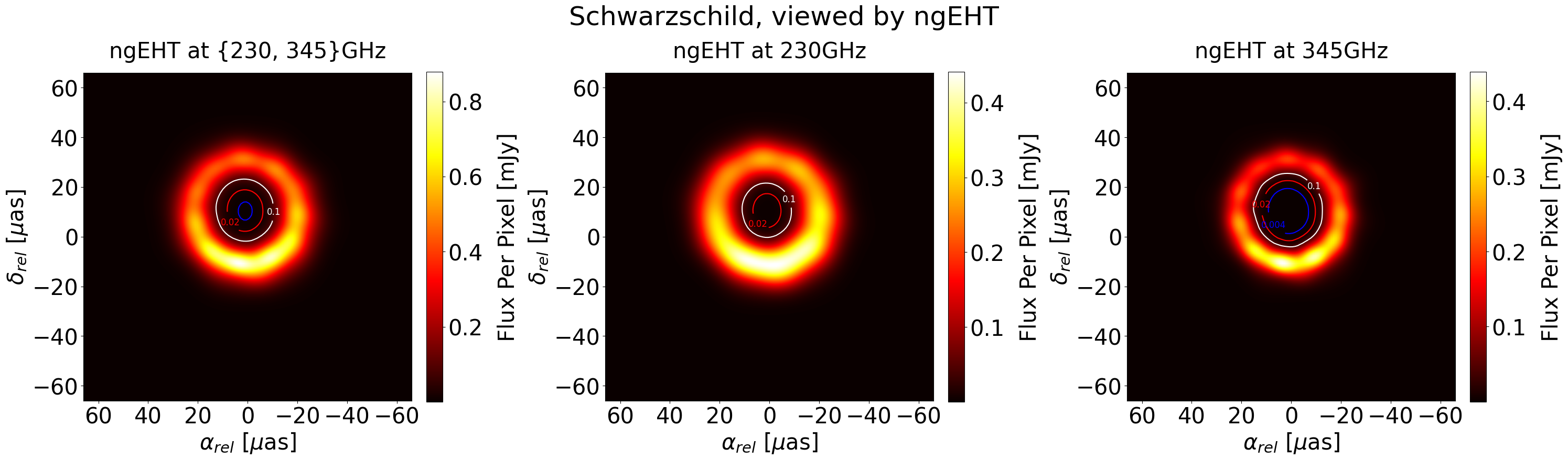}\\[3mm]
  \includegraphics[width=1.\textwidth]{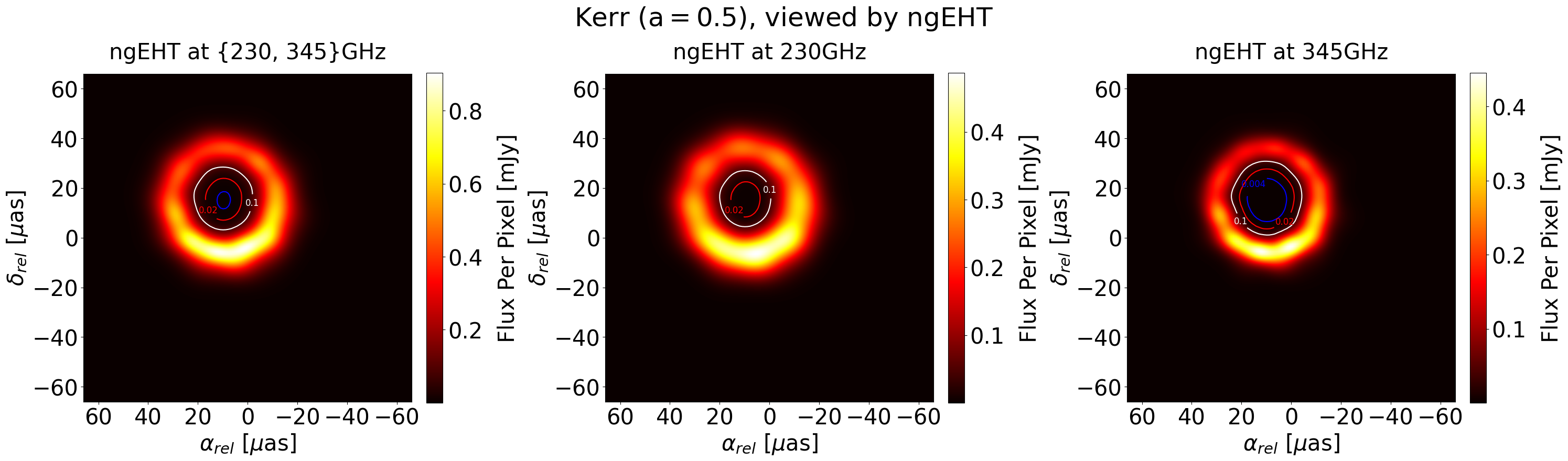}\\[3mm]
   \includegraphics[width=1.\textwidth]{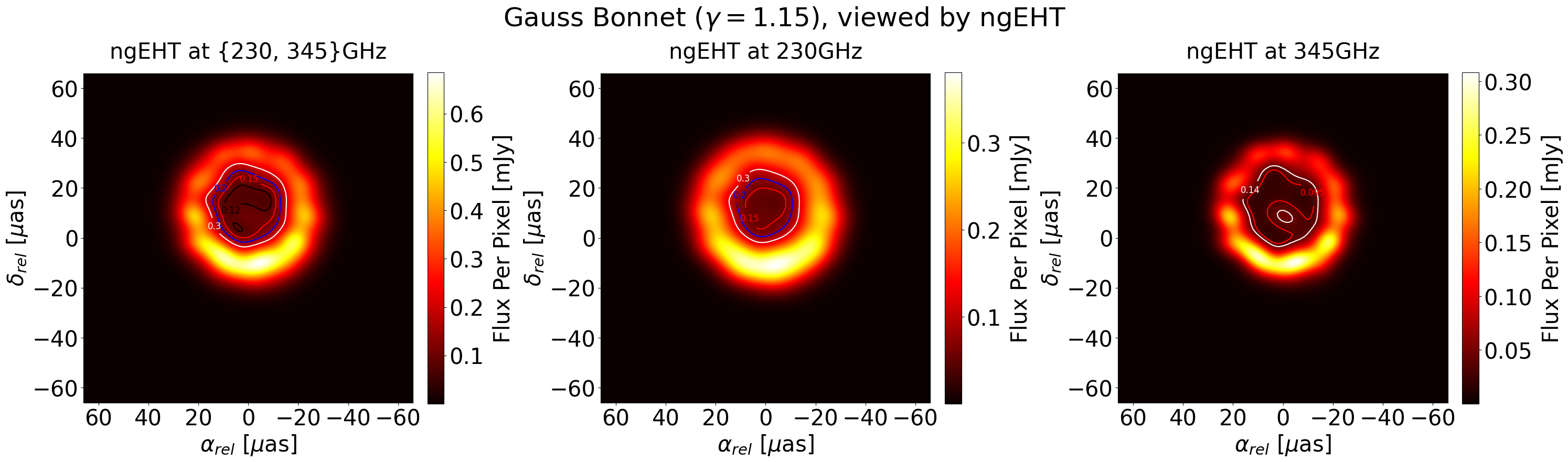}\\[3mm]
  \includegraphics[width=1.\textwidth]{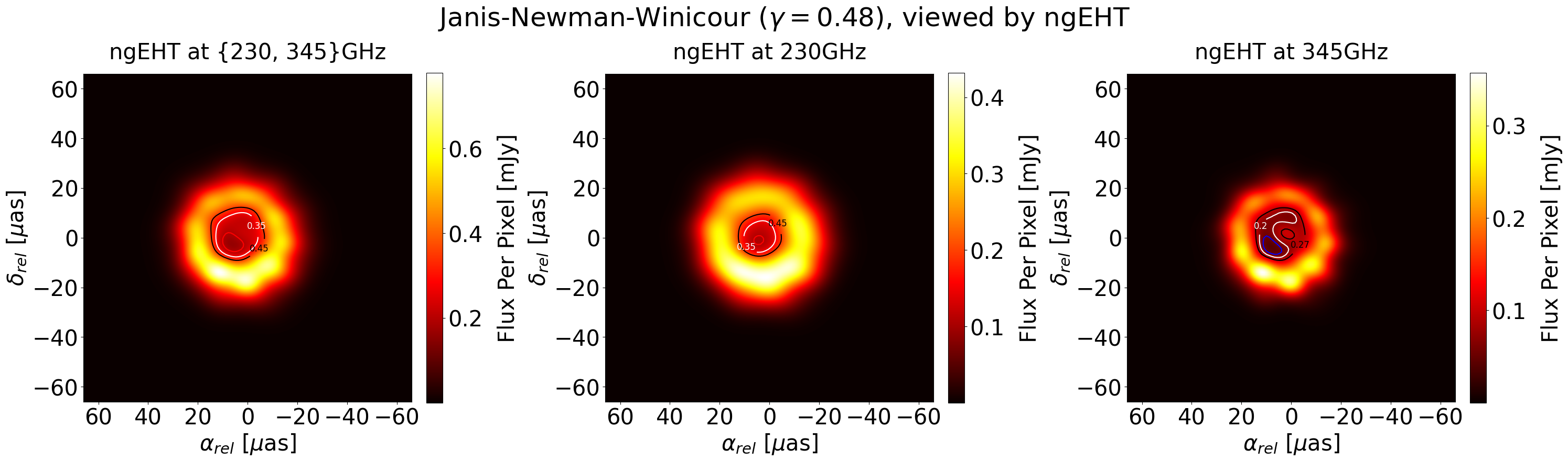}\\[1mm]
 \caption{Iso-flux contours at the central brightness depression of the reconstructed images as seen by the ngEHT. We consider two observing frequencies $\nu=230$ GHz and $\nu=345$ GHz and their superposition (left-most panel). The value of the flux at each contour is normalized to the maximal flux from the disk image.}
\label{fig:contour_ngEHT}
\end{figure}

We further notice that introducing a second frequency increases the resolution sufficiently so that the inner structure at the central part of the disk image becomes observable. The reconstructed images at the observing frequency $\nu=345$ GHz for the Gauss-Bonnet and JNW naked singularities contain bright regions in the central depression (see Figs. $\ref{fig:ehtim_ngEHT_NS}$ and $\ref{fig:ehtim_ngEHT_1}$). In order to evaluate the intensity of the central bright spots  we calculate the iso-flux contours of the central emission in Fig. $\ref{fig:contour_ngEHT}$. The values on each contour are normalized to the maximum flux from the disk image. We see that for the Gauss-Bonnet naked singularity at 345 GHz the bright central spots emit with approximately $15\%$ of the maximum intensity of the disk while for the Janis-Newman-Winicour solution the intensity reaches  $30\%$. The emission from the central structure is also evaluated in the VIDA template analysis in Fig. $\ref{fig:VIDA_ngEHT_11}$ by means of the intensity variation across the horizontal and vertical cross-sections through the center of the fitted ring template. This radiation is observationally significant showing that the next-generation EHT arrays will be capable of distinguishing qualitatively reflective naked singularities from black holes.


\section{Conclusion}

In this work we consider the observational properties of certain classes of naked singularities as seen by the current and near-future Even Horizon Telescope arrays. Naked singularities which possess a photon sphere frequently mimic the phenomenological  behavior of black holes. However, certain types of naked singularities act effectively as a reflective barrier for null geodesics causing the in-falling photon trajectories to scatter back to infinity. These spacetimes lead to clear-cut observational signatures in the morphology of the accretion disk images producing a series of bright rings at the central part of the image.

Although the central ring structure is a significant theoretical prediction, it is not clear how well it can be resolved with the current capacity of the Event Horizon Telescope. We explore this issue considering two particular cases of reflective naked singularities represented by the Janis-Newman-Winicour spacetime and a static spherically symmetric solution within the 4D Einstein-Gauss-Bonnet gravity. Using these geometries for modeling the galactic target M87$^*$ we demonstrate that the 2017 EHT arrays is incapable of resolving the central ring structure. We observe a central brightness depression in the disk images similar to the image structure for the  Kerr black hole. However, the presence of the bright rings leads to an increased intensity in the central part of the image compared to the Kerr black hole. The minimum flux in the central depression is with an order of magnitude higher than that for the Kerr black hole providing a quantitative measure for distinguishing the reflective naked singularity spacetimes.

The upgraded next-generation Event Horizon Telescope arrays significantly increase the capacity of observing the central ring structure. Considering observations at 230 GHz we obtain that the deviation of the central brightness depression intensity from the Kerr black hole reaches two orders of magnitude. Moreover, introducing a second observational frequency at 345 GHz increases the resolution sufficiently so that we can distinguish a  distribution of bright spots at the central part of the image. Thus, the next-generation Event Horizon Telescope provides already the opportunity  for detecting quantitative effects produced by  reflective naked singularities by means of the modified morphology of their disk images.

\section*{Appendix}

\subsection*{A. Telescope arrays}

In this section we describe the specifications of the EHT telescope arrays which we use in order to reconstruct the observable disk images. We consider three telescope arrays including the arrays which were used in the 2017 and 2022 EHT observations and a tentative array suggested  for near-future observations.  The 2022 EHT array is designed by adding three telescopes to the 2017 EHT array observing at 230 GHz \cite{EHT2}, \cite{EHT_1}. The ngEHT array is an extension containing 13 more telescopes and supporting two observing frequencies at $\nu = 230$ GHz and $\nu= 345$ GHz \cite{ngEHT}. In Table $\ref{tab:telescopes}$ we provide the locations of the telescope stations included in each array as well as the system equivalent flux density (SEFD) for each station at 230 GHz and 345 GHz.

\begin{table}
\begin{flushleft}
\footnotesize
	\begin{tabular}{l||r|r|r||r|r}
		2017 EHT array & $X_\text{geo}$ [m] & $Y_\text{geo}$ [m] & $Z_\text{geo}$ [m] & $\text{\tiny SEFD}_{230}$ [Jy] & $\text{\tiny SEFD}_{345}$ [Jy]
		\\\hline\hline
		PV &
		$5088968$ &
		$-301682$ &
		$3825016$ &
		$330$&
		$3850$
		\\
		AZ &
		$-1828796$ &
		$-5054407$ &
		$3427865$ &
		$2850$ &
		$17190$
		\\
		SM &
		$-5464523$ &
		$-2493147$ &
		$2150612$ &
		$1230$ &
		$5730$
		\\
		LM &
		$-768714$ &
		$-5988542$ &
		$2063276$ &
		$110$ &
		$2040$
		\\
		AA &
		$2225061$ &
		$-5440057$ &
		$-2481681$ &
		$40$ &
		$250$
		\\
		SP &
		$0$ &
		$0$ &
		$-6359610$ &
		$7510$ &
		$25440$
		\\
		AP &
		$2225040$ &
		$-5441198$ &
		$-2479303$ &
		$1790$ &
		$8880$
		\\
		JC &
		$-5464585$ &
		$-2493001$ &
		$2150654$ &
		$1190$ &
		$5780$
		\\\hline\hline
            2022 EHT array \\ \hline\hline
		GL &
		$541647$ &
		$-1388536$ &
		$6180829$ &
		$4350$ &
		$14390$
		\\
		PB &
		$4523999$ &
		$468045$ &
		$4460310$ &
		$300$ &
		$1410$
		\\
		KP &
		$-1994314$ &
		$-5037909$ &
		$3357619$ &
		$7430$ &
		$44970$
		\\\hline\hline
            ngEHT array \\ \hline\hline
		BA &
		$-2352576$ &
		$-4940331$ &
		$3271508$ &
		$16930$&
		$58500$
		\\
		BR &
		$-2363000$ &
		$-4445000$ &
		$3907000$ &
		$15770$ &
		$52160$
		\\
		CI &
		$5311000$ &
		$-1725000$ &
		$3075000$ &
		$19410$ &
		$76110$
		\\
		GB &
		$5627890$ &
		$ 1637767$ &
		$-2512493$ &
		$14270$ &
		$264200$
		\\
		OV &
		$-2409598$ &
		$-4478348$ &
		$3838607$ &
		$15100$ &
		$118890$
		\\
		SG &
		$1832000$ &
		$-5034000$ &
		$-3455000$ &
		$17570$ &
		$63760$
		\\
		CT &
		$1569000$ &
		$-4559000$ &
		$-4163000$ &
		$29580$ &
		$167890$
		\\
		GR &
		$1538000$ &
		$-2462000$ &
		$-5659000$ &
		$71080$ &
		$736630$
		\\
		HA &
		$1521000$ &
		$-4417000$ &
		$4327000$ &
		$2740$ &
		$66530$
		\\
		NZ &
		$-4540000$ &
		$719000$ &
		$-4409000$ &
		$32040$ &
		$191080$
	\end{tabular}
	\caption{\label{tab:telescopes}
	Location and system equivalent flux density (SEFD) at $230$ GHz and $345$ GHz for the telescope stations included in the EHT arrays. The 2017 EHT array consists of the first 8 telescopes, while the 2022 EHT array includes in addition the next three stations. The ngEHT array is extended with the last 10 stations including all the telescopes in the list.}
\end{flushleft}
\end{table}

\subsection*{B. Simulations and image reconstruction for near-future arrays}

\begin{figure}[h!]
\centering
  \includegraphics[width=0.75\textwidth]{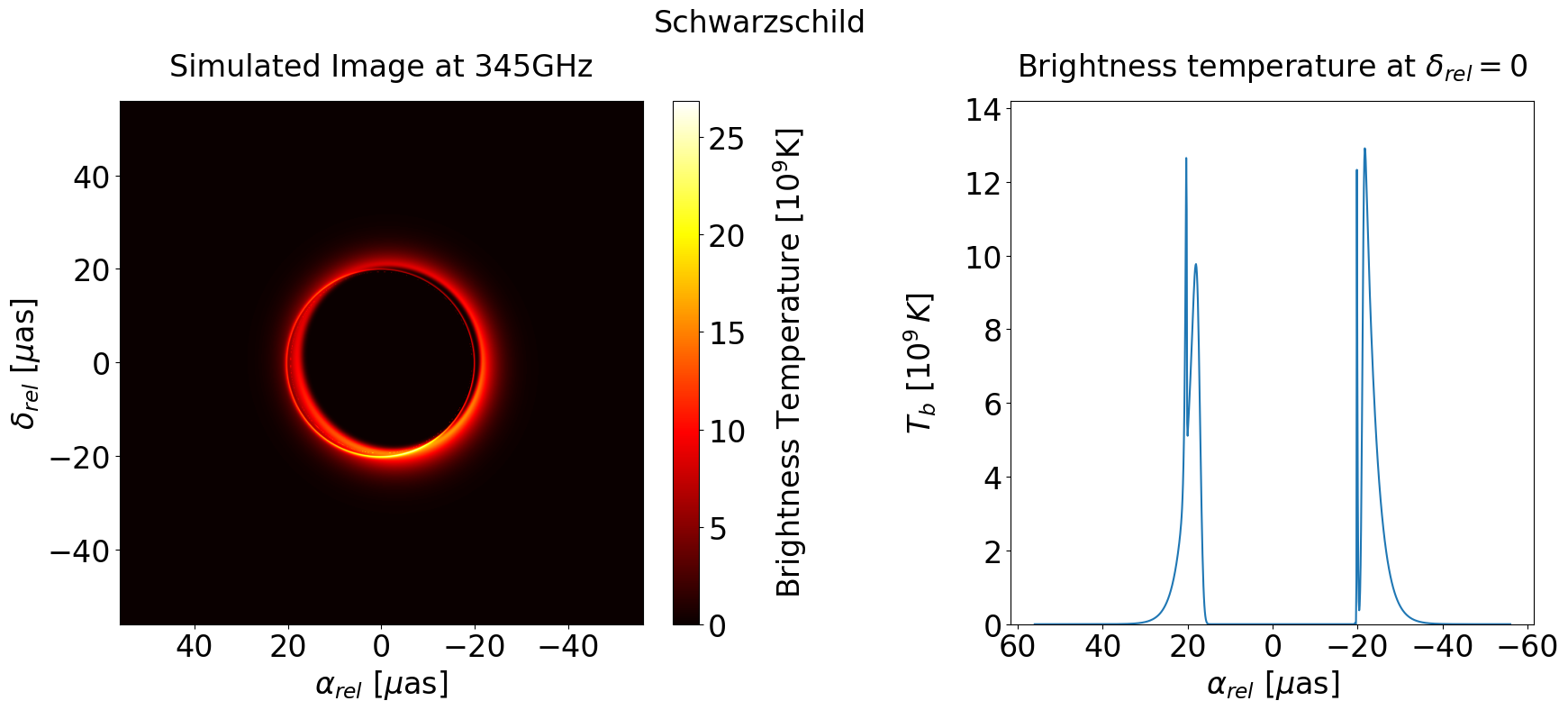}\\
  \includegraphics[width=0.75\textwidth]{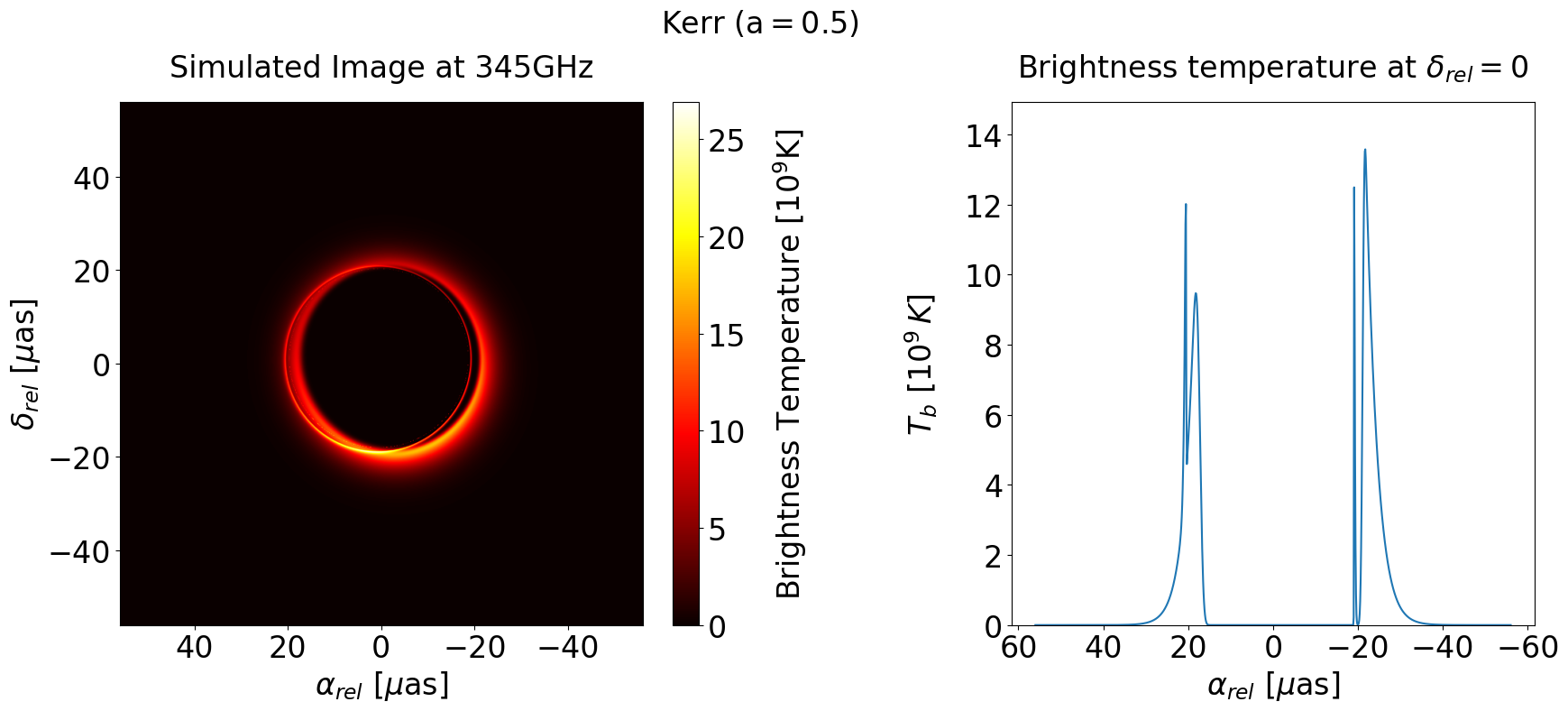} \\
   \includegraphics[width=0.75\textwidth]{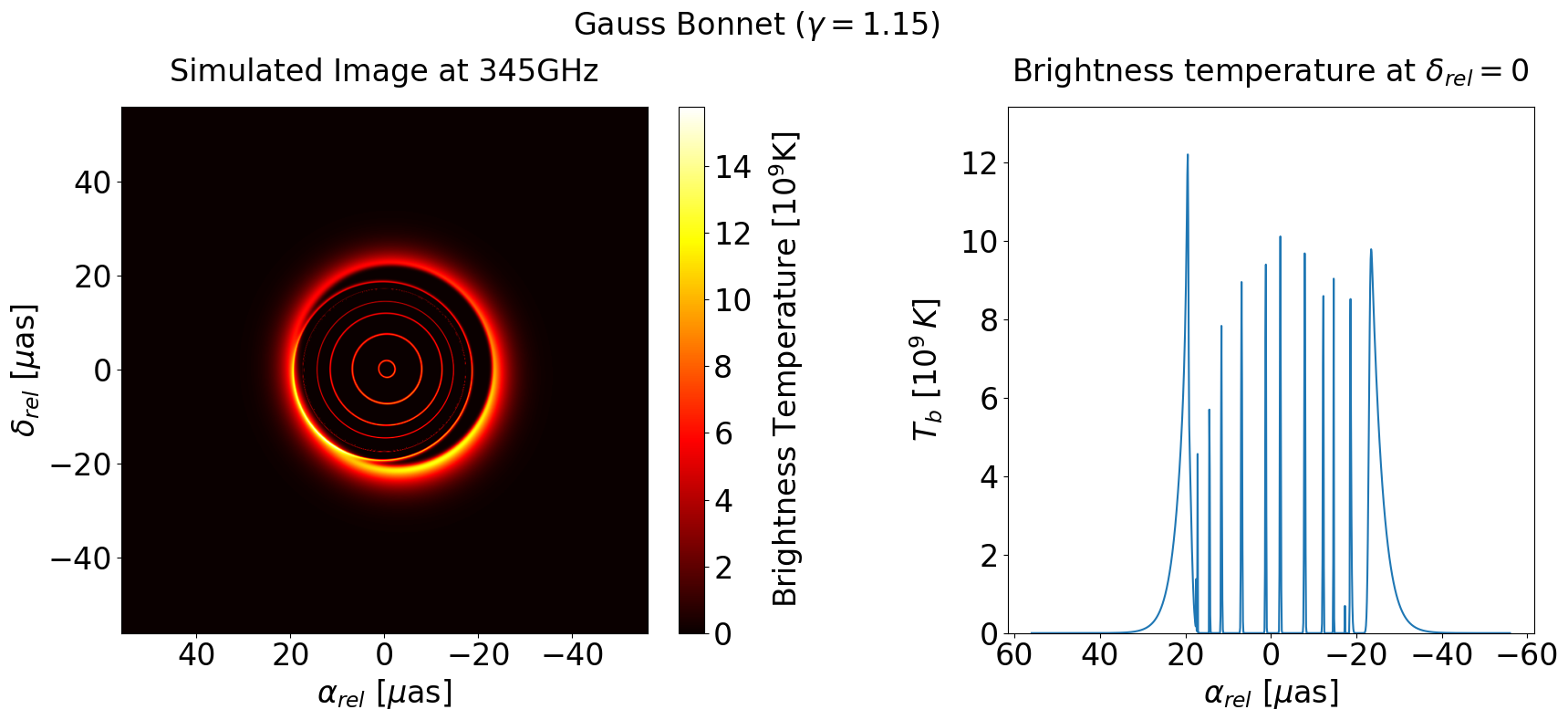}\\
  \includegraphics[width=0.75\textwidth]{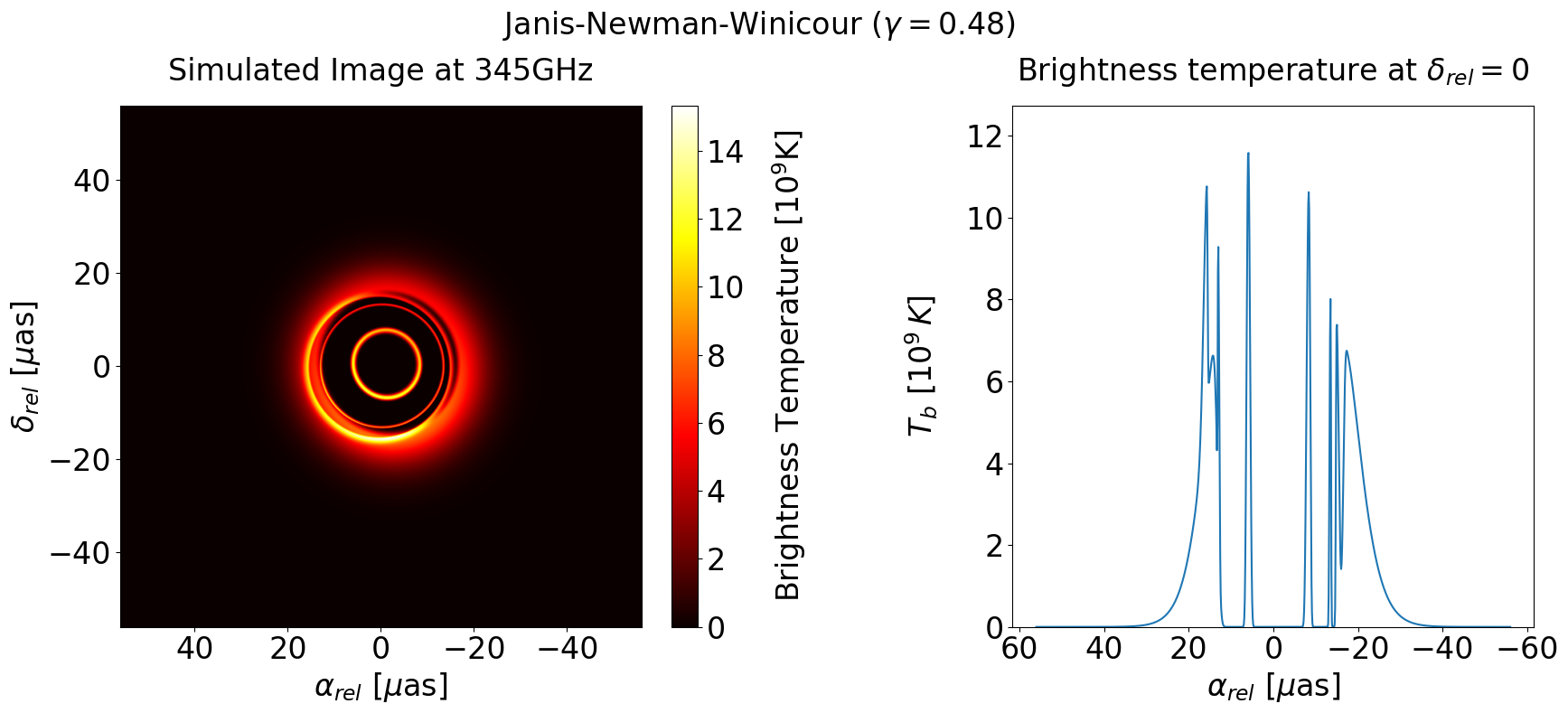}
 \caption{Simulated images at the observing frequency $\nu=345$ GHz.  For the  parameters of the accretion disk model and the ray tracing procedure see Tables $\ref{table:Ray_tr}$ and $\ref{table:param_sim}$. }
\label{fig:Ray_tr_345}
\end{figure}

In this section we perform a numerical ray-tracing in the Gauss-Bonnet and Janis-Newman-Winicour spacetimes in order to simulate the theoretically observable intensity from the accretion disk at the observing frequency $\nu= 345$ GHz. The images are compared to the Schwarzschild black hole and the Kerr black hole with a spin parameter $a=0.5$ (see Fig. $\ref{fig:Ray_tr_345}$). We follow  the conventions introduced in sections 3 and 4 considering the phenomenological RIAF disk model described in Eq. ($\ref{disk_model}$) emitting synchrotron radiation. The images are adapted to the observational settings of the galactic target M87* with parameters of the ray-tracing procedure and disk model given in Tables $\ref{table:Ray_tr}$ and $\ref{table:param_sim}$.

We further obtain the reconstructed images of the accretion disks in the naked singularities and black hole spacetimes as seen by two near-future EHT arrays offering better spacial resolution (see Appendix A). The images are presented in Figs. $\ref{fig:ehtim_2022}$-$\ref{fig:ehtim_ngEHT_1}$ following the conventions introduced in section 4 while the goodness-of-fit parameters associated with the visibility amplitudes $\chi^2_{\text{vis}}$ and the closure phases $\chi^2_{\text{cl}}$ in each reconstruction are given in Table $\ref{table:chi_2022}$.

\bigskip

\begin{figure}[h!]
\centering
  \includegraphics[width=0.75\textwidth]{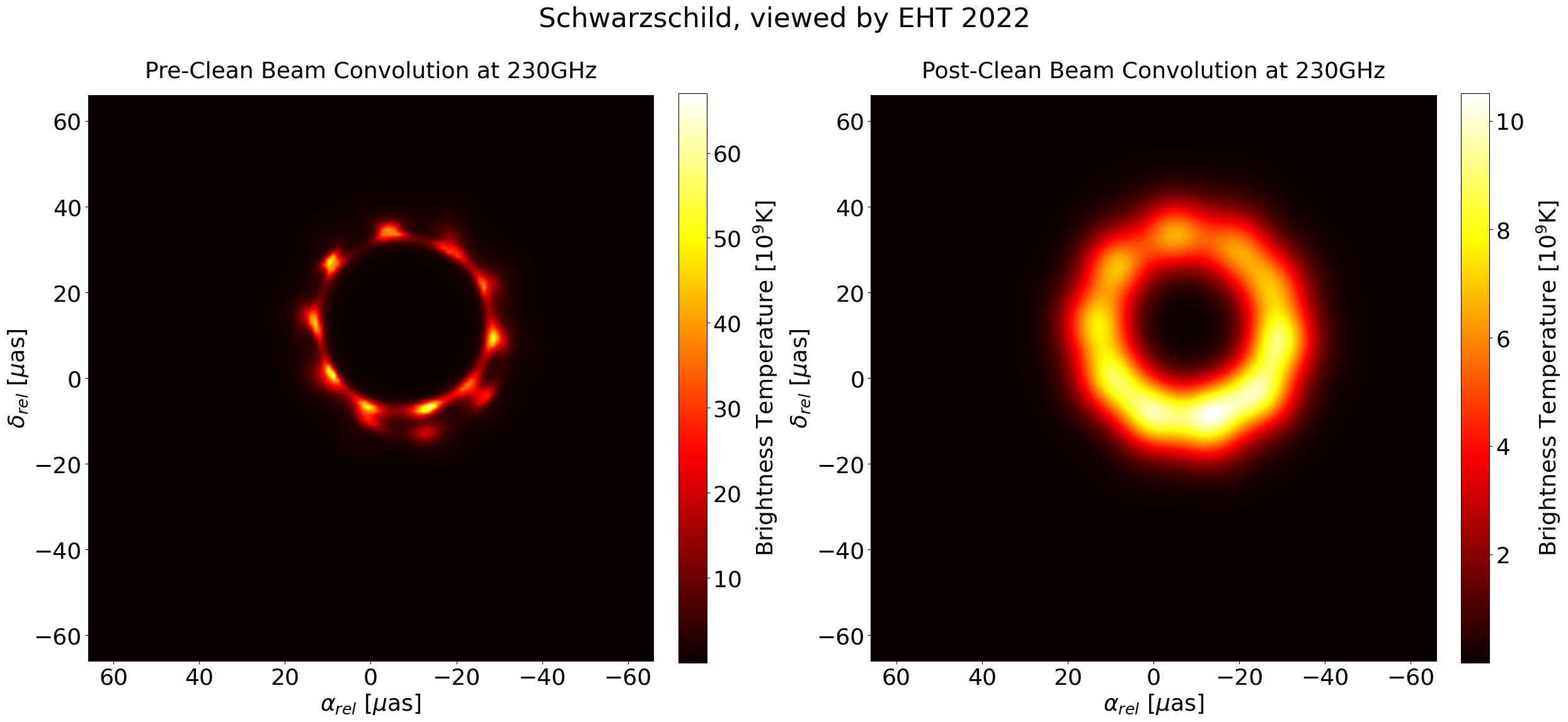}\\
  \includegraphics[width=0.75\textwidth]{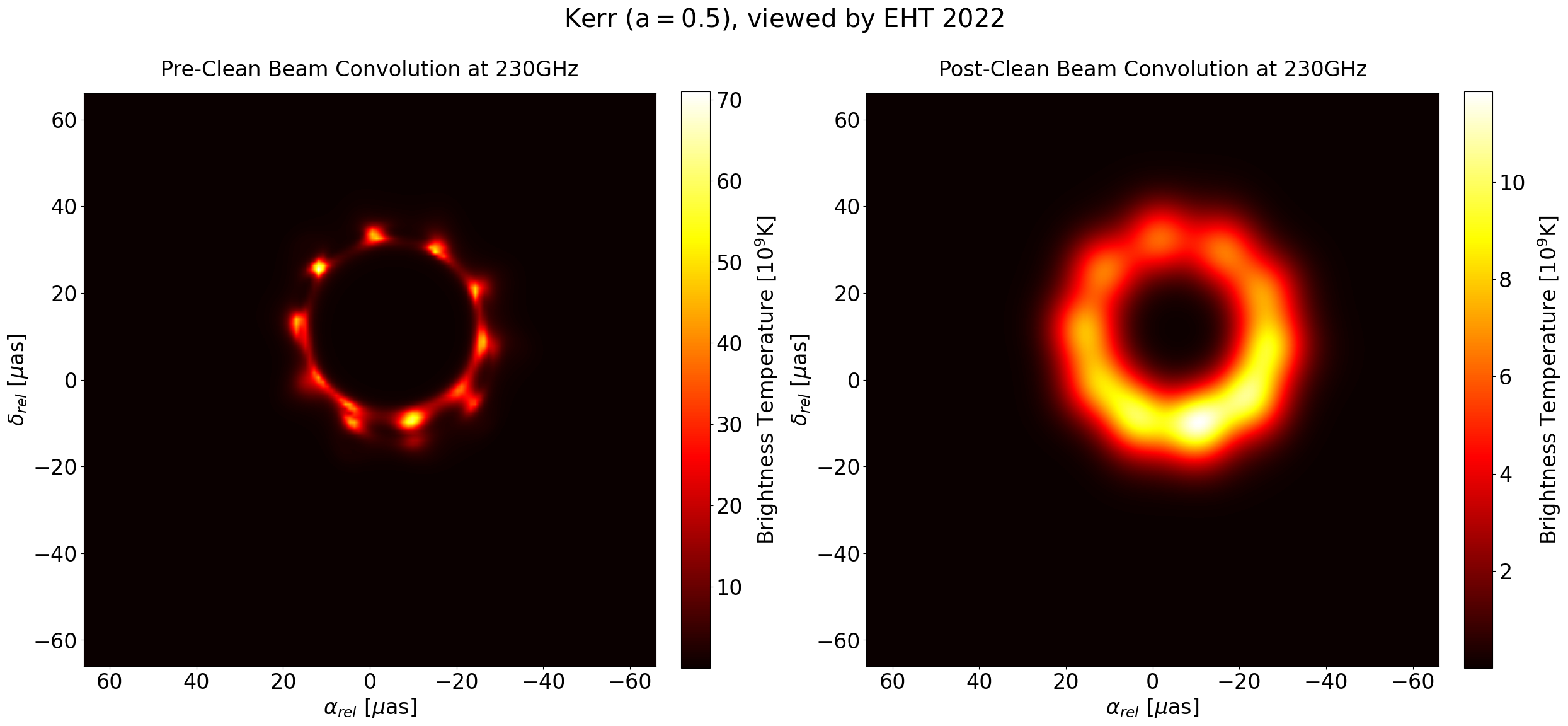} \\
  \includegraphics[width=0.75\textwidth]{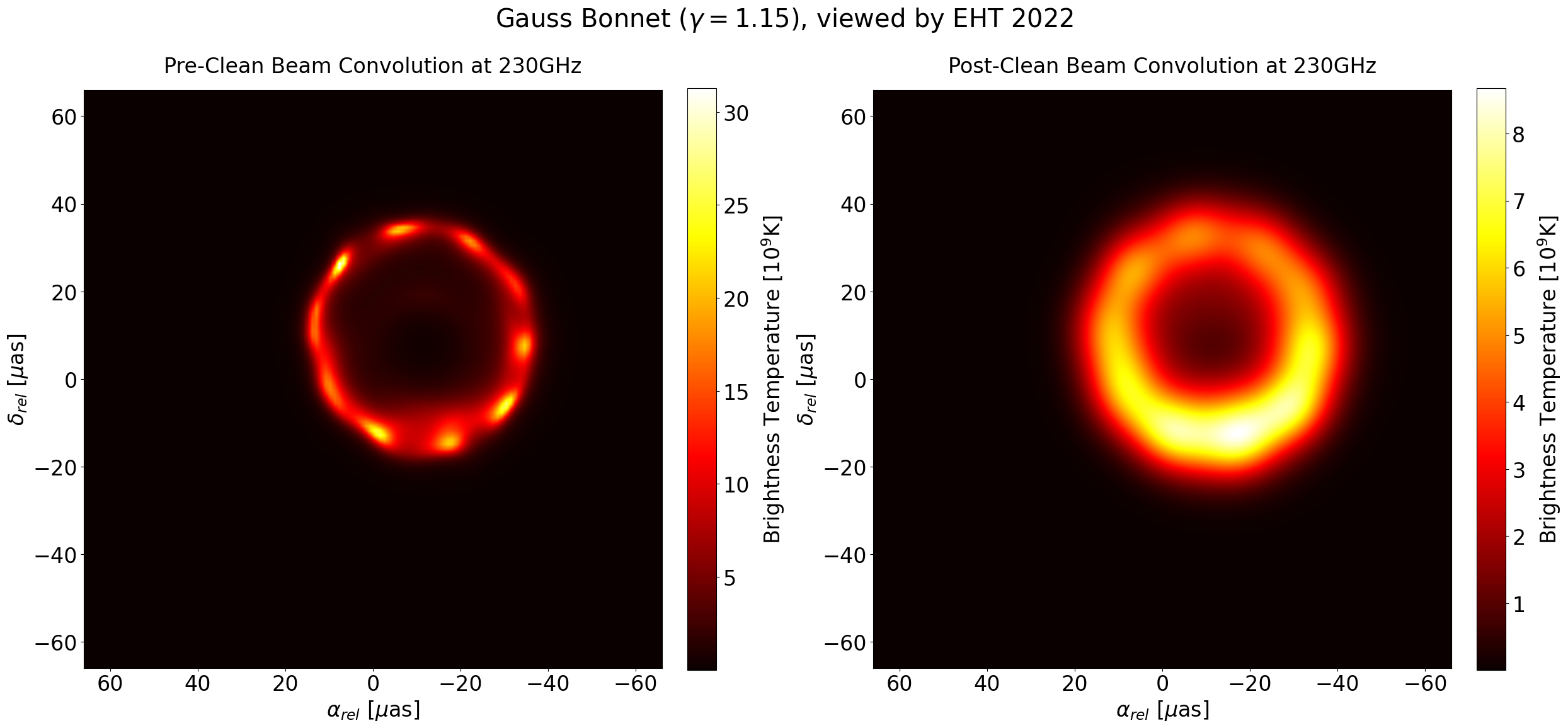}\\
  \includegraphics[width=0.75\textwidth]{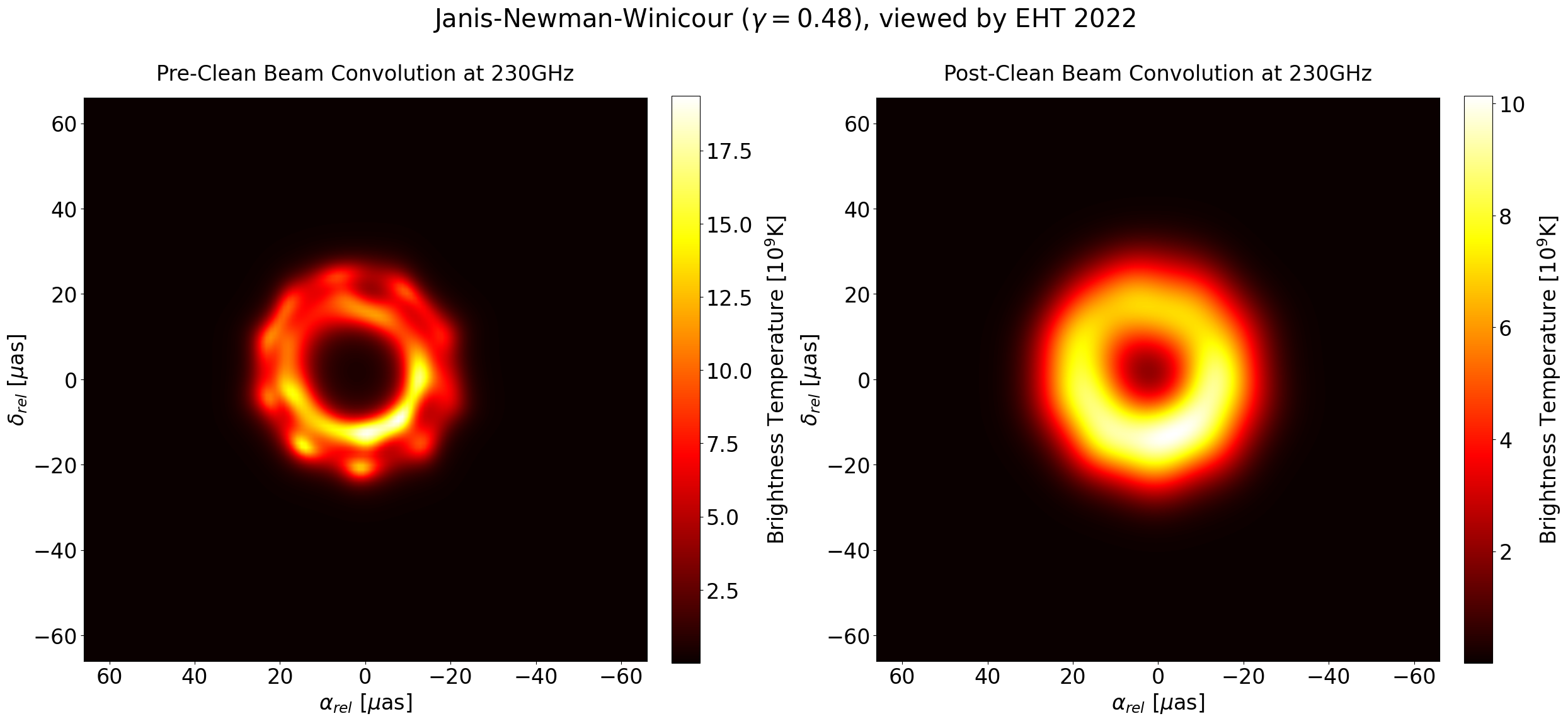}
 \caption{Reconstructed images as seen by the 2025 EHT array. The left panel provides the image at full resolution while the image on the right is blurred with $1/2$ interferometer clean beam. The goodness-of-fit parameters are given in Table $\ref{table:chi_2022}$.}
\label{fig:ehtim_2022}
\end{figure}

\begin{figure}[h!]
\centering
  \includegraphics[width=0.75\textwidth]{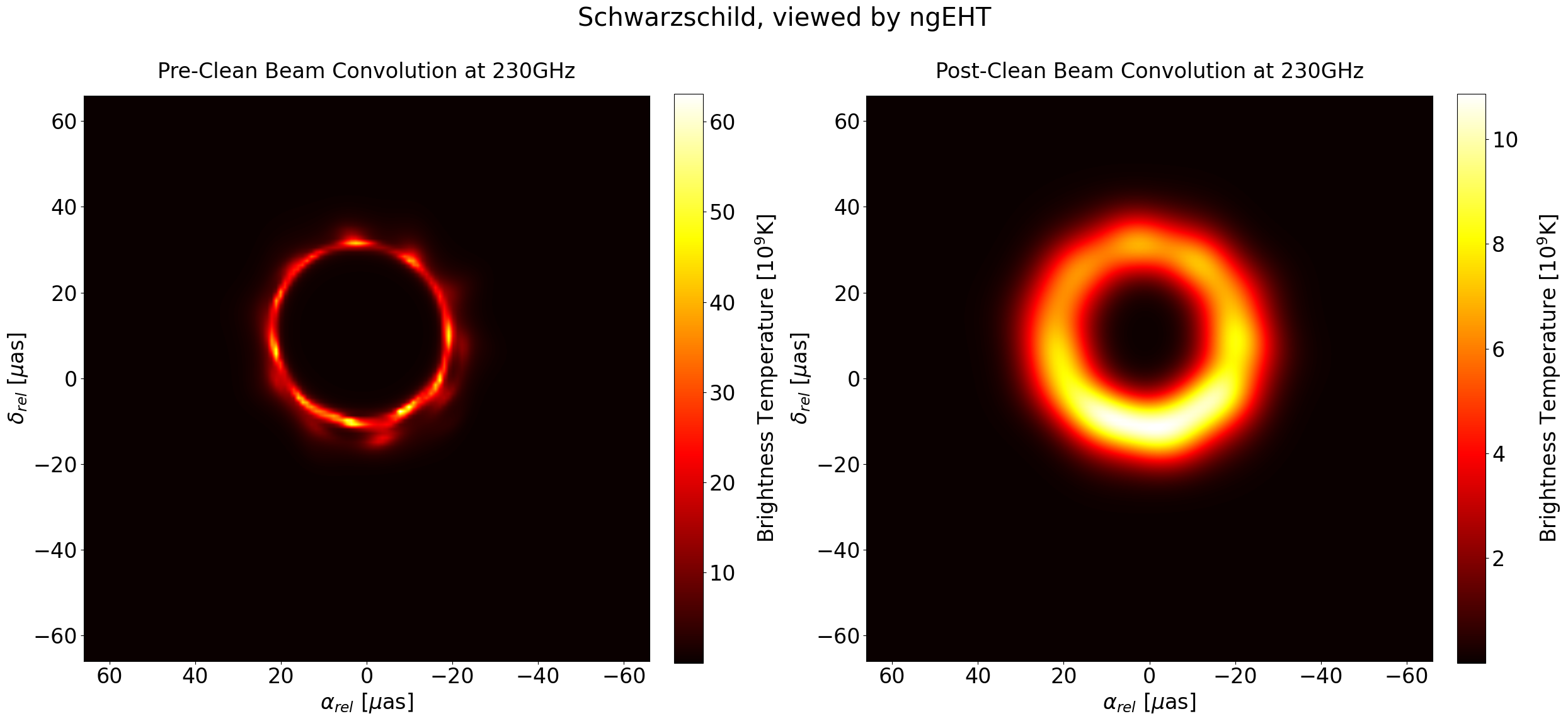}\\
  \includegraphics[width=0.75\textwidth]{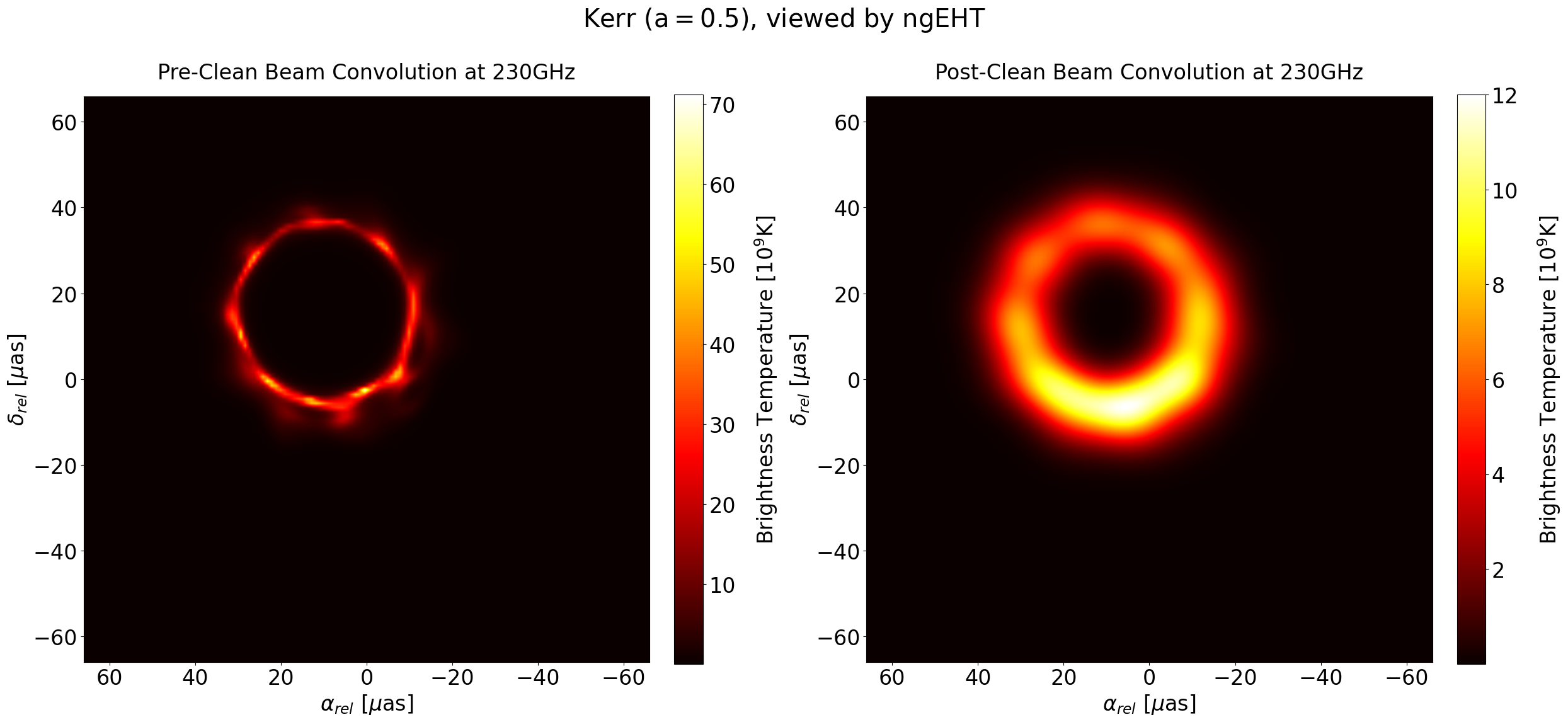}\\
  \includegraphics[width=0.75\textwidth]{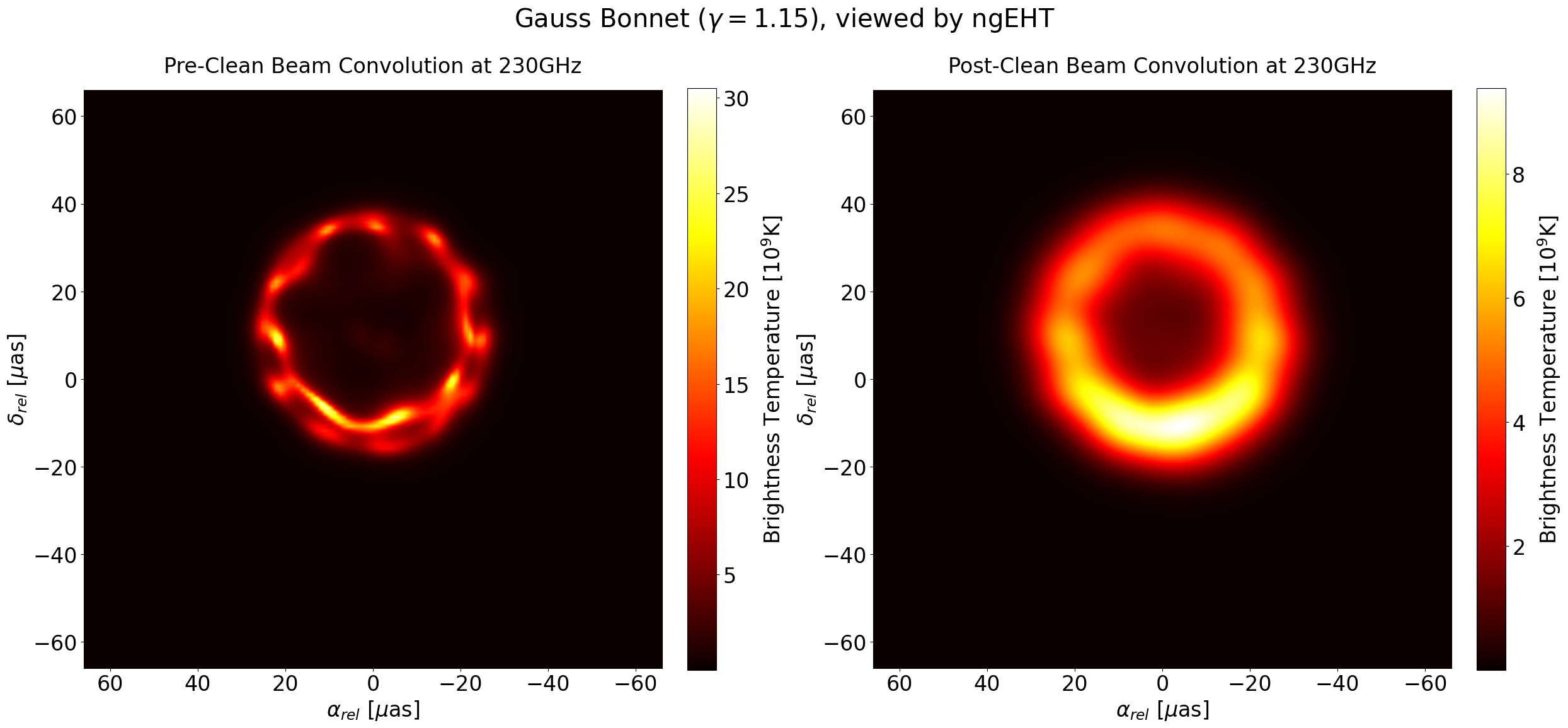}\\
  \includegraphics[width=0.75\textwidth]{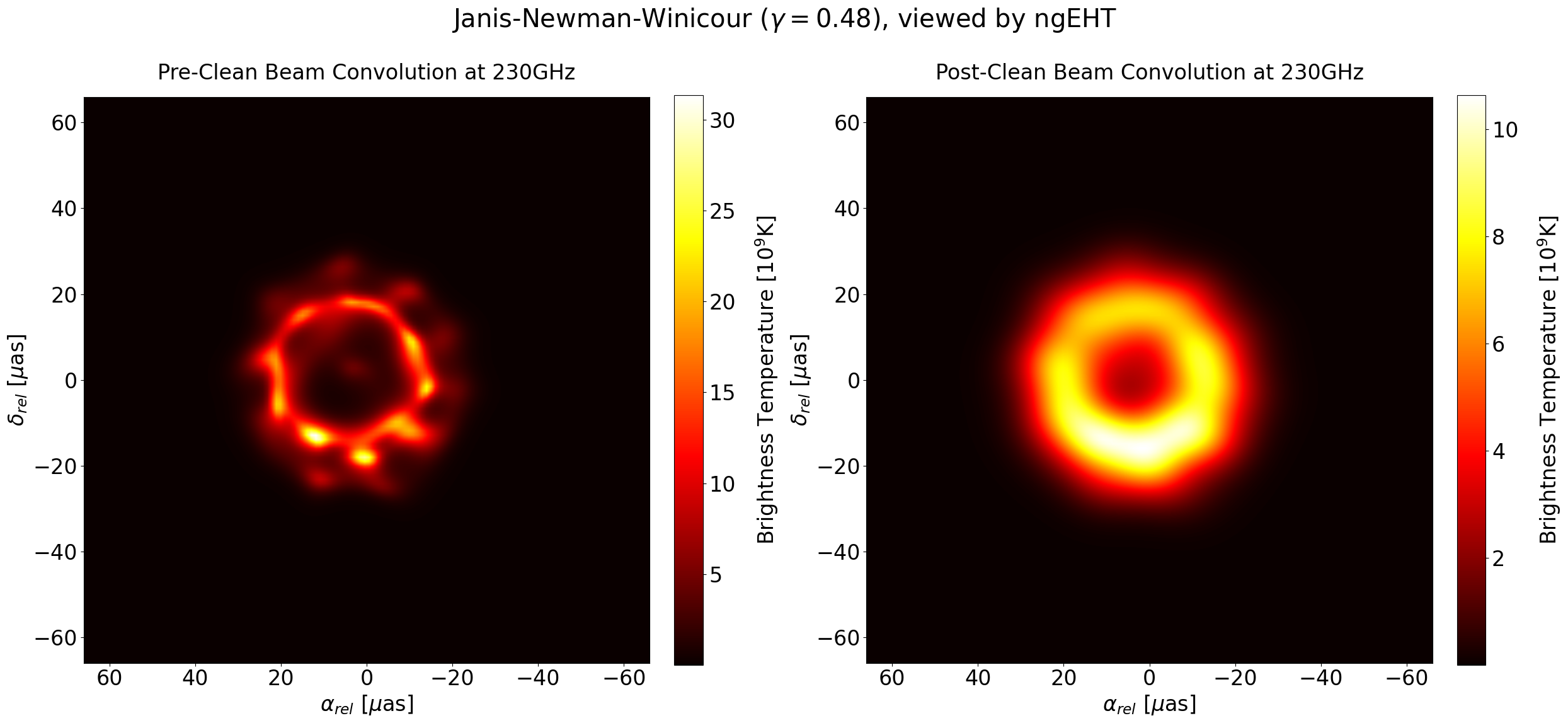}
 \caption{Reconstructed images as seen by the ngEHT array at the observing frequency $\nu=230$ GHz. The left panel provides the image at full resolution while the image on the right is blurred with $1/2$ interferometer clean beam. The goodness-of-fit parameters are given in Table $\ref{table:chi_2022}$.}
\label{fig:ehtim_ngEHT}
\end{figure}

\begin{figure}[h!]
\centering
  \includegraphics[width=0.75\textwidth]{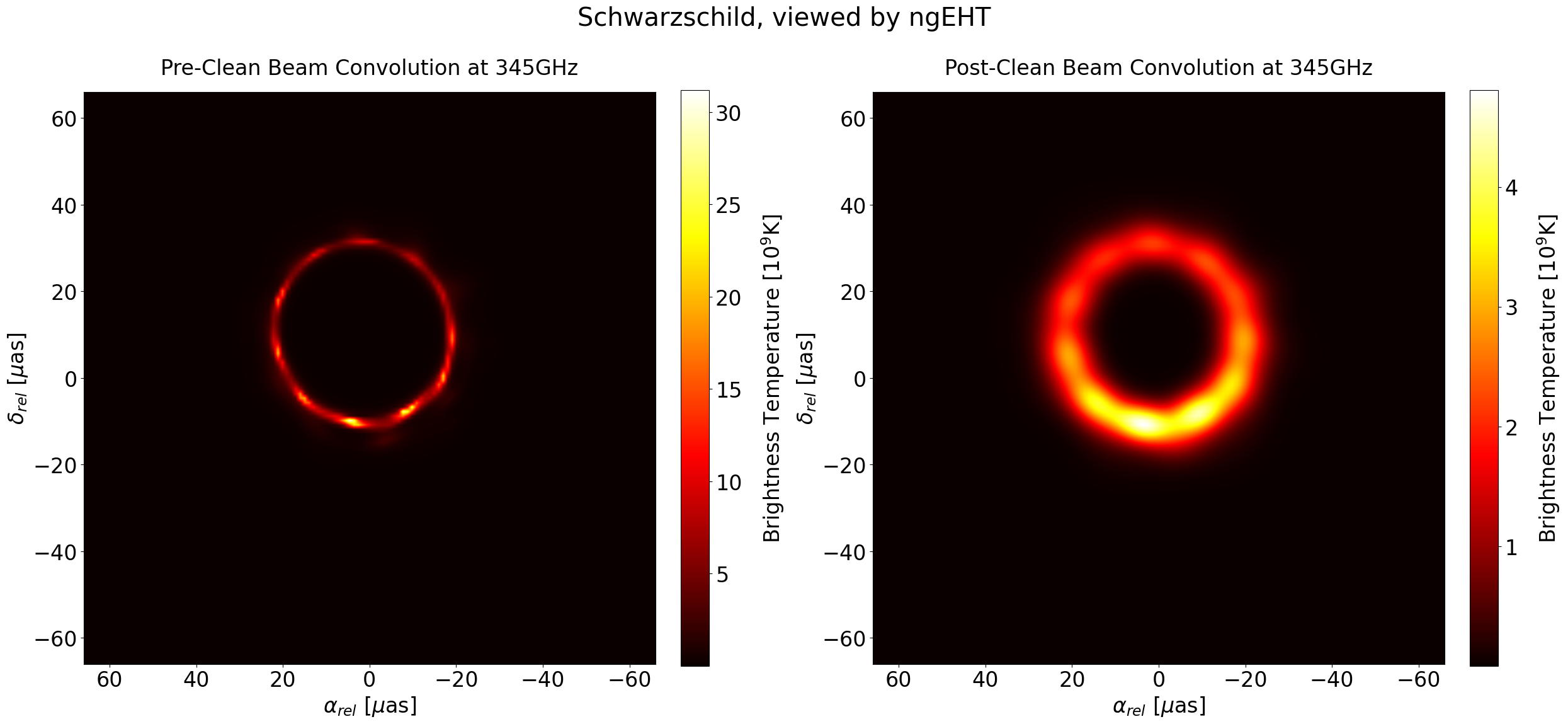}\\
  \includegraphics[width=0.75\textwidth]{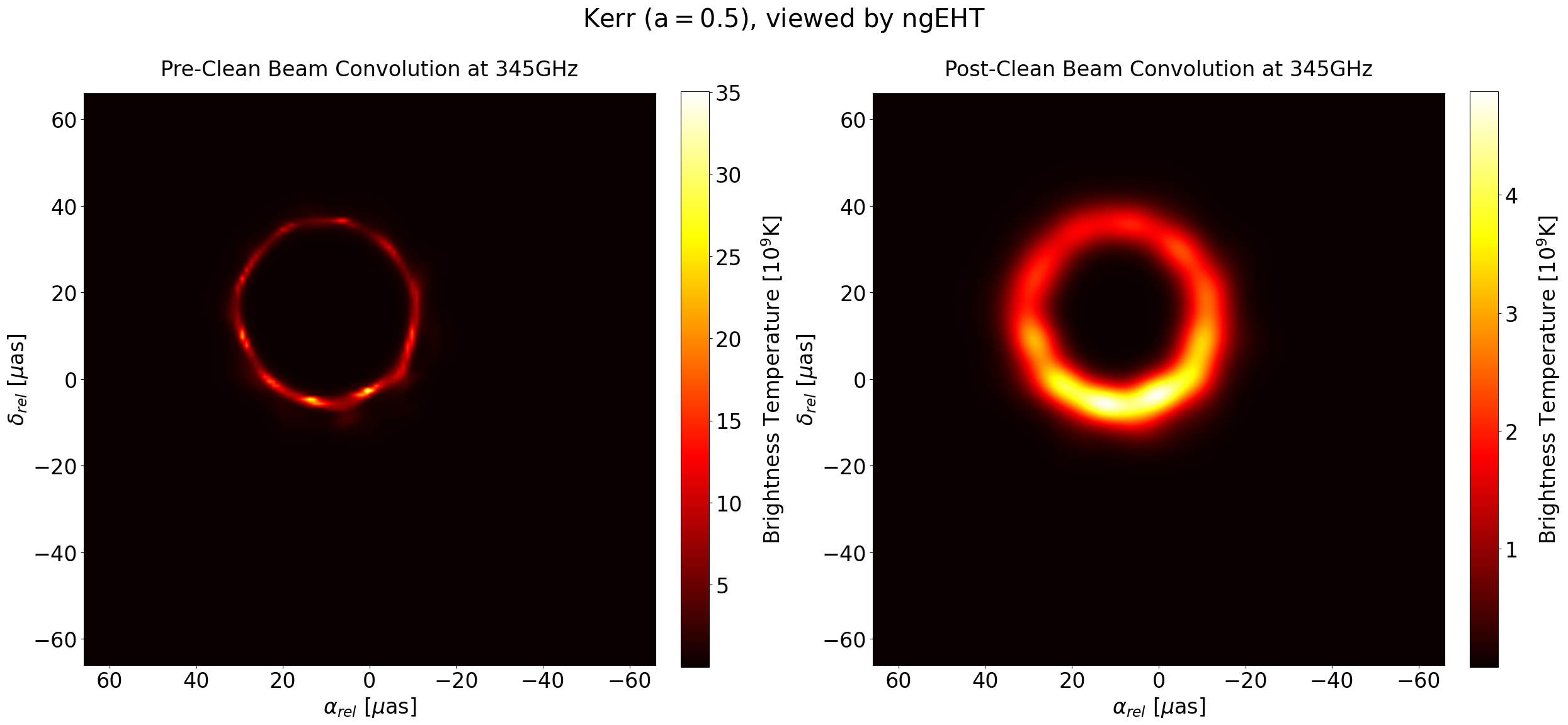} \\
  \includegraphics[width=0.75\textwidth]{GB_ehtim_ngEHT_1.png}\\
  \includegraphics[width=0.75\textwidth]{JNW_ehtim_ngEHT_1.png}
 \caption{Reconstructed images as seen by the ngEHT array at the observing frequency $\nu=345$ GHz. The left panel provides the image at full resolution while the image on the right is blurred with $1/2$ interferometer clean beam. The goodness-of-fit parameters are given in Table $\ref{table:chi_2022}$.}
\label{fig:ehtim_ngEHT_1}
\end{figure}

\begin{table}[h!]
\centering
\begin{tabular}{c|c|c|c|c}
            \hline
		{spacetime} & {Schw.}&{\thead{Kerr \\ (a=0.5)}}&{Gauss-Bonnet}&{JNW}
		\\\hline\hline
		\thead{$r_0$ [M] \\ (cut-off radius)} & 4.5& 4.5& 5& 5
		\\\hline
     \thead{ $T_0$ [$\times \, 10^{10}$ K] \\ (inner temperature at $r_0$)} &6.8 & 6.8 & 5.9 & 7.2 
		\\\hline
      		\thead{ $F$ [Jy] \\ (total flux)} & 0.30& 0.31& 0.26& 0.29
		\\\hline
 	\end{tabular}
\caption{Parameters of the simulated accretion disk images at the observing frequency $\nu=345$ GHz. }
\label{table:param_sim}
\end{table}

\begin{table}[h!]
\centering
\begin{tabular}{c|c|c|c|c}
            \hline
		{spacetime} & {Schw.}&{\thead{Kerr \\ (a=0.5)}}&{Gauss-Bonnet}&{JNW}
		\\\hline\hline
		\thead{$\chi^2_{\text{vis}}$ \\ (EHT 2022)} & 1.09& 1.14& 1.05 & 1.08
		\\\hline
     \thead{ $\chi^2_{\text{cl}}$ \\ (EHT 2022)} & 1.05& 0.98& 0.98 & 1.01
		\\\hline\hline
      		\thead{$\chi^2_{\text{vis}}$ \\ (ngEHT 230 GHz)} & 1.04&1.04&1.04&1.00
		\\\hline
  \thead{$\chi^2_{\text{cl}}$ \\ (ngEHT 230 GHz)}& 1.02&0.99&1.06&0.99
		\\\hline\hline
		\thead{$\chi^2_{\text{vis}}$ \\ (ngEHT 345 GHz)} & 0.98&0.99&0.99&1.00
		\\\hline
  \thead{$\chi^2_{\text{cl}}$ \\ (ngEHT 345 GHz)}& 1.47&1.40&1.45&1.53
		\\\hline
	\end{tabular}
\caption{Goodness-of-fit parameters associated with the visibility amplitudes $\chi^2_{\text{vis}}$ and the closure phases $\chi^2_{\text{cl}}$ for the reconstructed images as seen by the 2022 EHT array and the ngEHT array at the observing frequencies $\nu=230$ GHz and $\nu=345$ GHz. }
\label{table:chi_2022}
\end{table}

\bigskip

\subsection*{C. VIDA template analysis for near-future arrays}

We perform a geometrical analysis on the reconstructed images presented in Appendix B fitting the accretion disk images to a Gaussian ring template. For the purpose we use the VIDA toolkit as the procedure is explained in detail in section 4. The model is parameterized by the same parameters and their values in each particular case are presented in Tables $\ref{table:VIDA_2022}$-\ref{table:VIDA_superposition}.

\begin{figure}[h!]
\centering
  \includegraphics[width=\textwidth]{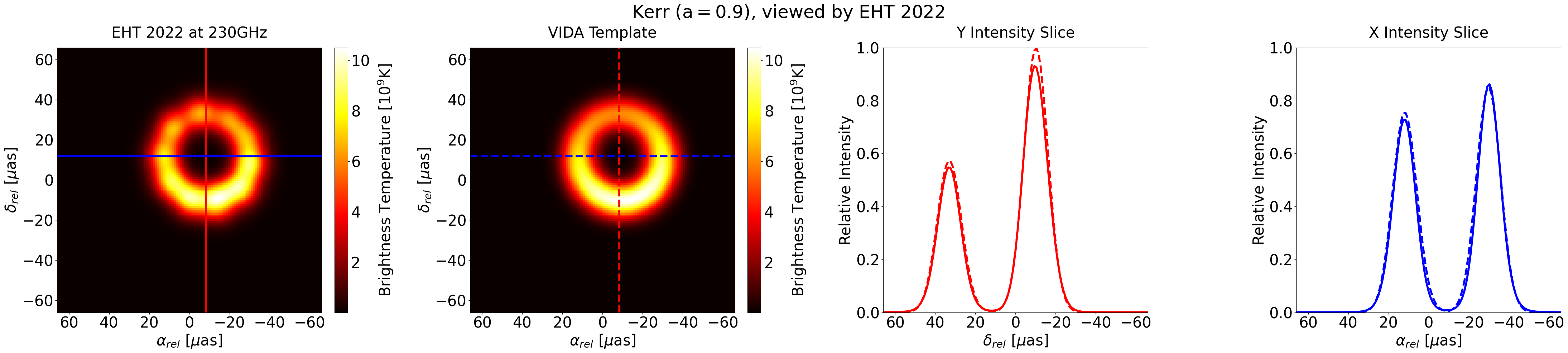}\\[3mm]
  \includegraphics[width=\textwidth]{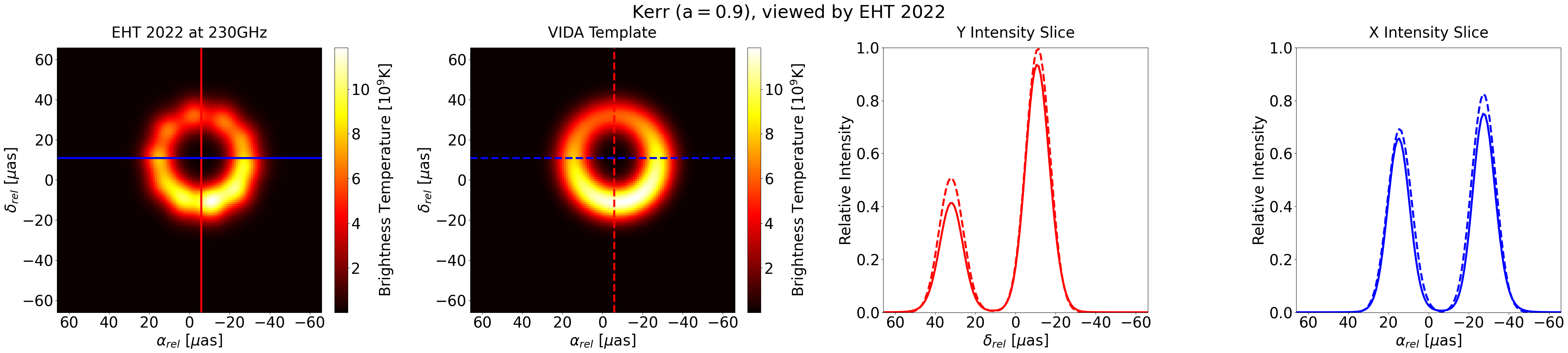}\\[3mm]
  \includegraphics[width=\textwidth]{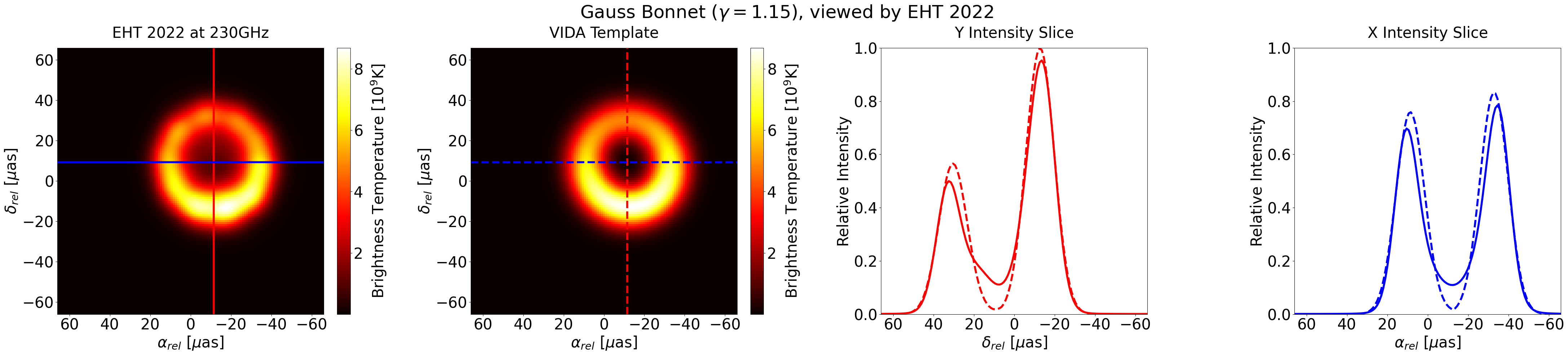}\\[3mm]
   \includegraphics[width=\textwidth]{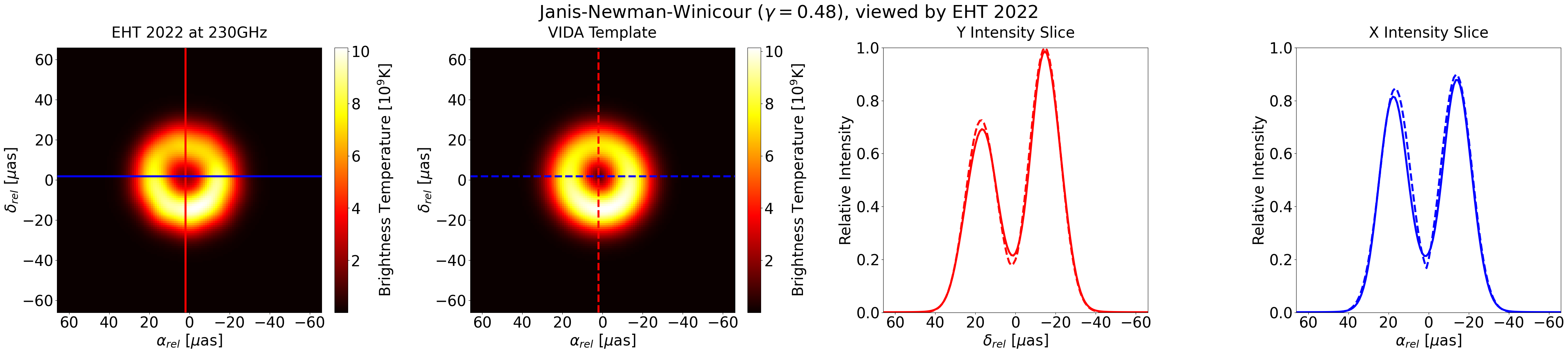}
 \caption{Reconstructed images (left) and VIDA fit with a ring template (right) for the accretion disks as seen by the 2022 EHT array.  In the right panel we investigate the  variation of the radiation intensity across the horizontal  (blue line) and  vertical (red line) cross-section through the center of the fitted ellipse. The intensity cross-sections are compared for the reconstructed images (solid lines) and the fitted VIDA templates (dashed lines). }
\label{fig:VIDA_2022}
\end{figure}

\begin{figure}[h!]
\centering
  \includegraphics[width=\textwidth]{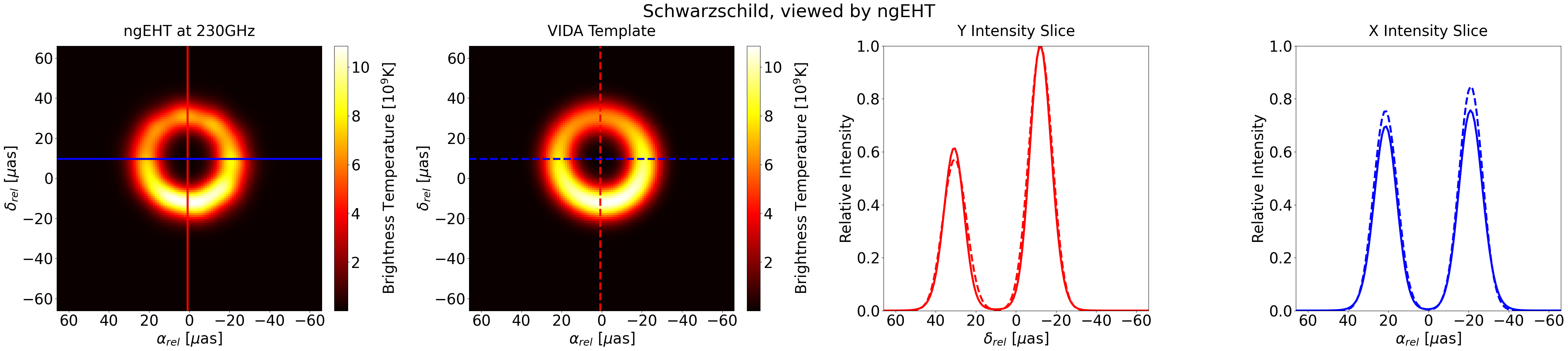}\\[3mm]
  \includegraphics[width=\textwidth]{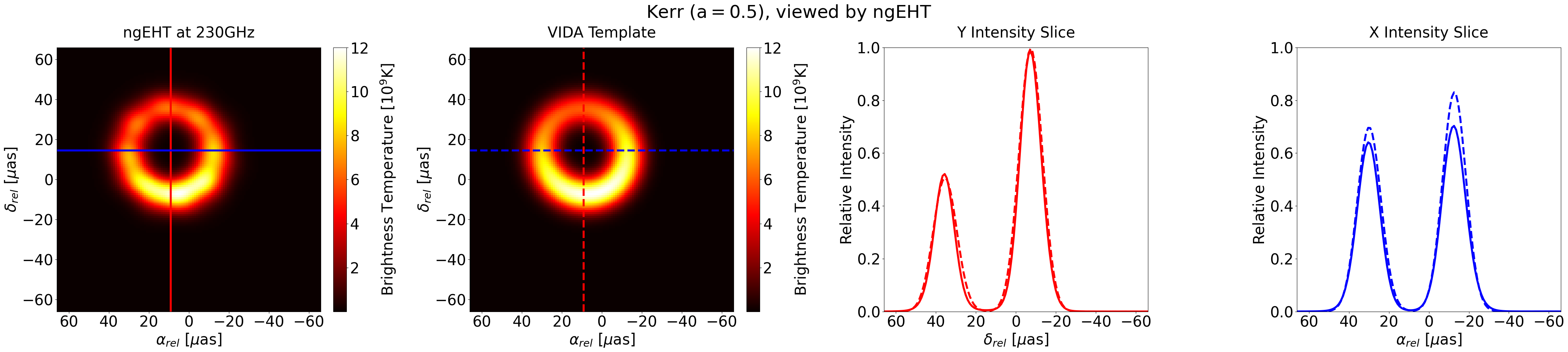}\\[3mm]
  \includegraphics[width=\textwidth]{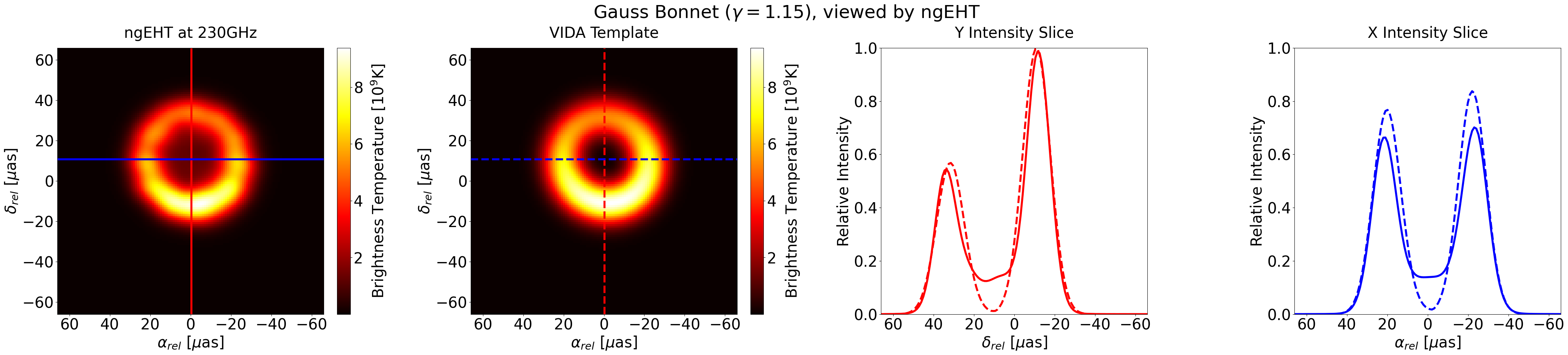}\\[3mm]
   \includegraphics[width=\textwidth]{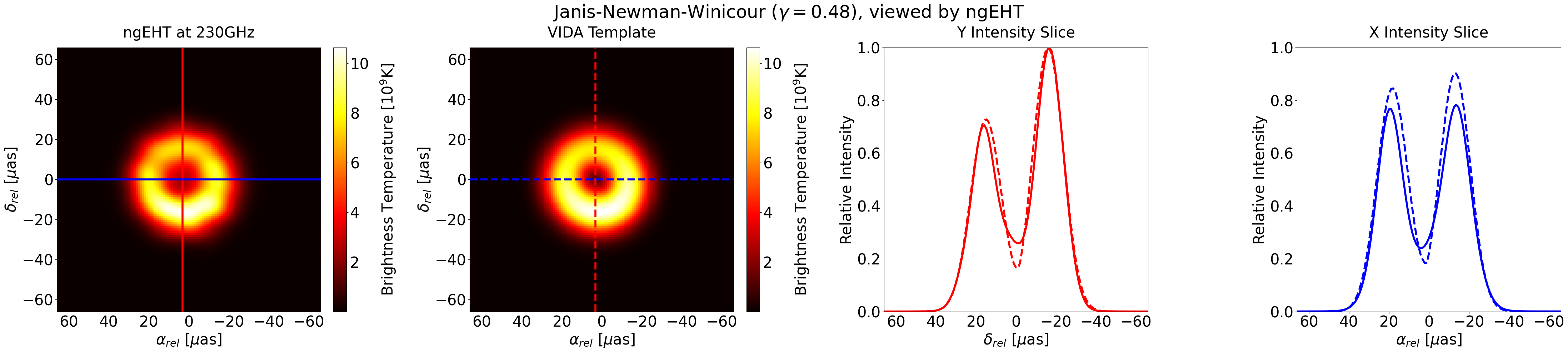}
 \caption{Reconstructed images (left) and VIDA fit with a ring template (right) for the accretion disks as seen by the ngEHT array at the observing frequency $\nu=230$ GHz.  In the right panel we investigate the  variation of the radiation intensity across the horizontal  (blue line) and  vertical (red line) cross-section through the center of the fitted ellipse. The intensity cross-sections are compared for the reconstructed images (solid lines) and the fitted VIDA templates (dashed lines). }
\label{fig:VIDA_ngEHT}
\end{figure}

\begin{figure}[h!]
\centering
  \includegraphics[width=\textwidth]{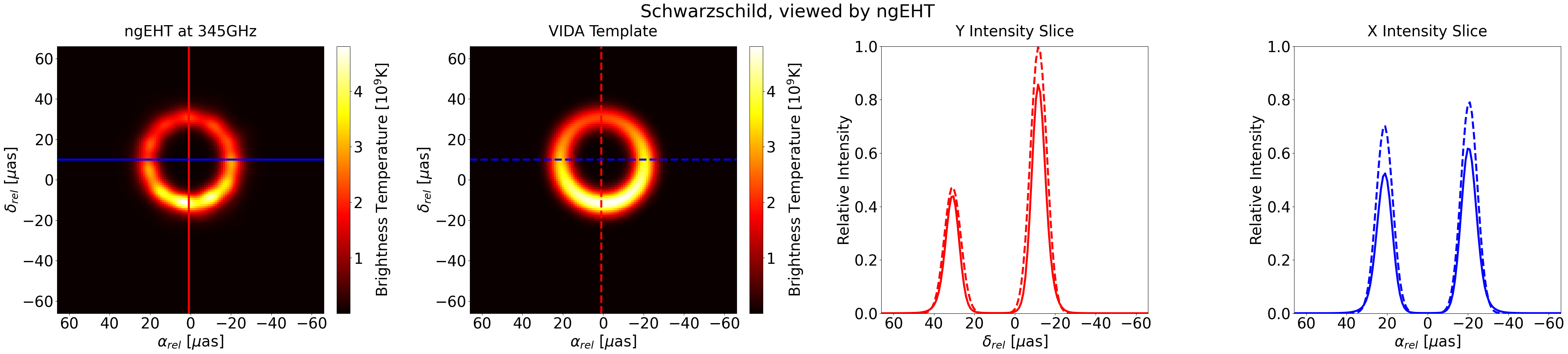}\\[3mm]
  \includegraphics[width=\textwidth]{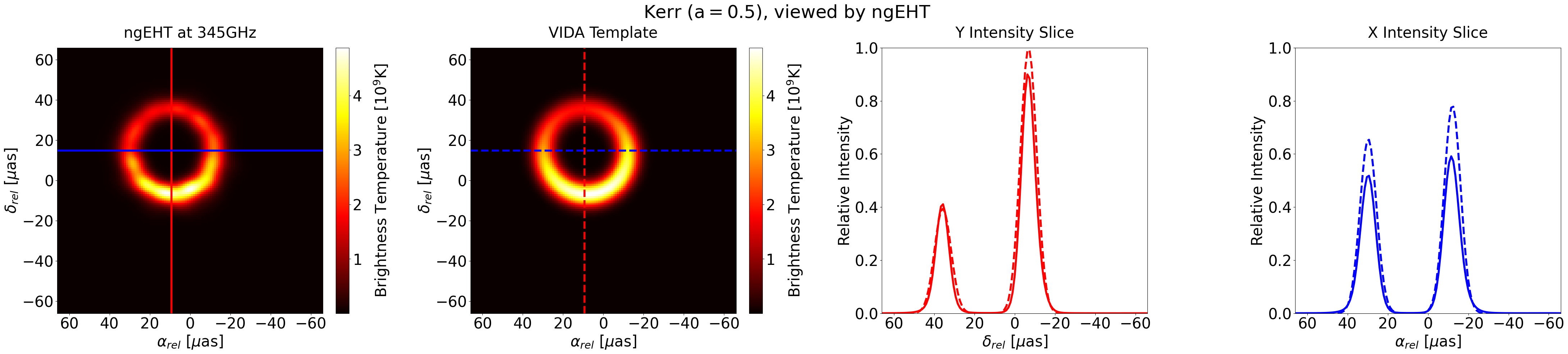}\\[3mm]
  \includegraphics[width=\textwidth]{GB_Vida_ngEHT_1.png}\\[3mm]
   \includegraphics[width=\textwidth]{JNW_Vida_ngEHT_1.png}
 \caption{Reconstructed images (left) and VIDA fit with a ring template (right) for the accretion disks as seen by the ngEHT array at the observing frequency $\nu=345$ GHz. In the right panel we investigate the  variation of the radiation intensity across the horizontal  (blue line) and  vertical (red line) cross-section through the center of the fitted ellipse. The intensity cross-sections are compared for the reconstructed images (solid lines) and the fitted VIDA templates (dashed lines). }
\label{fig:VIDA_ngEHT_1}
\end{figure}

\begin{figure}[h!]
\centering
  \includegraphics[width=\textwidth]{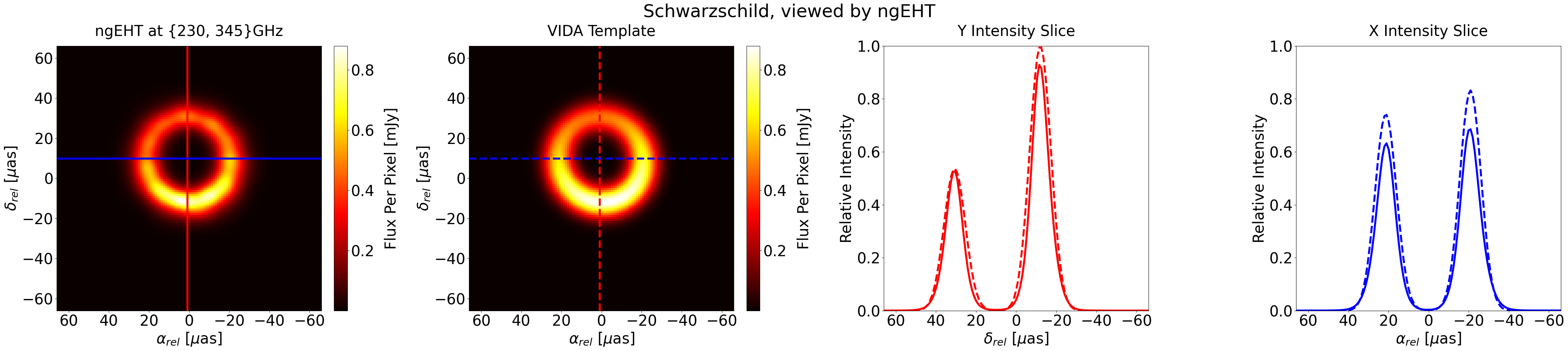}\\[3mm]
  \includegraphics[width=\textwidth]{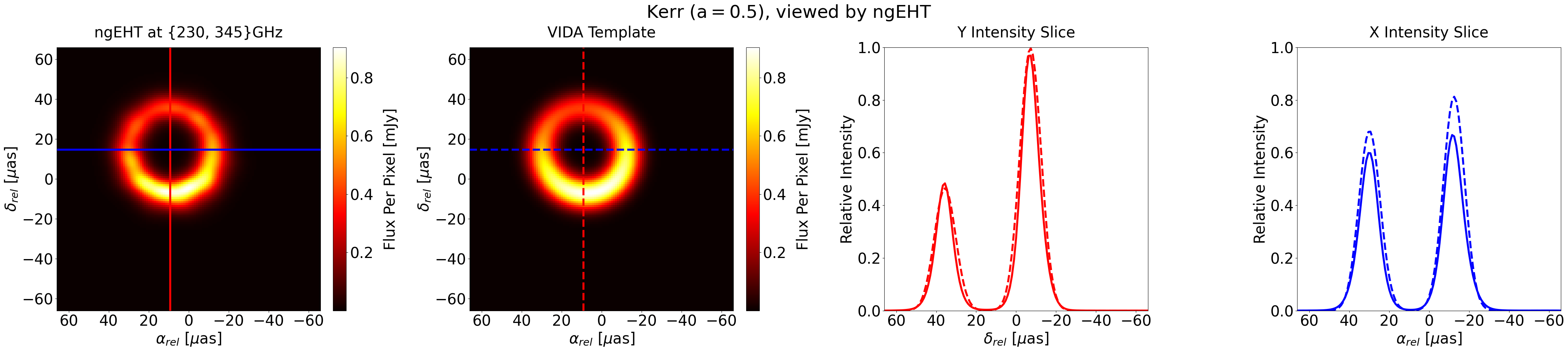}\\[3mm]
  \includegraphics[width=\textwidth]{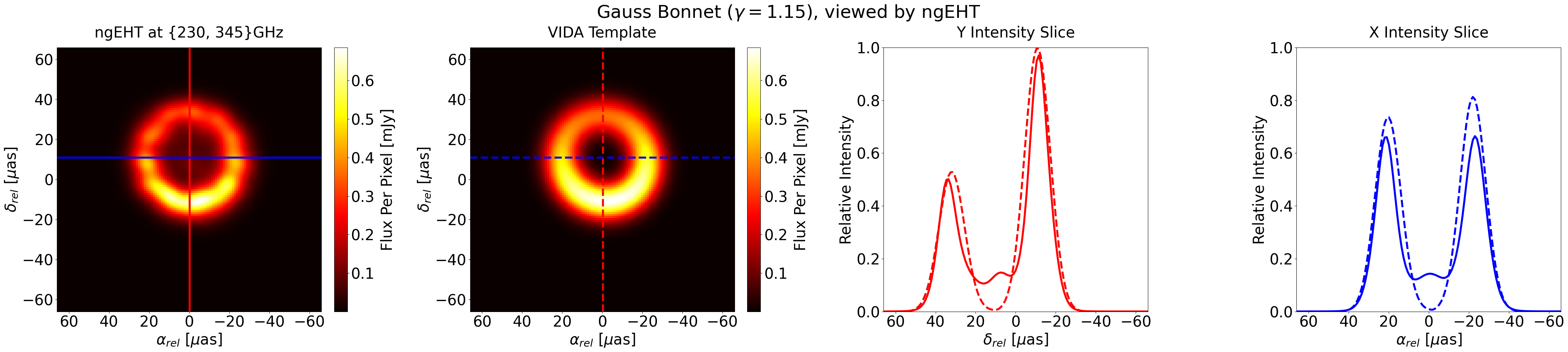}\\[3mm]
   \includegraphics[width=\textwidth]{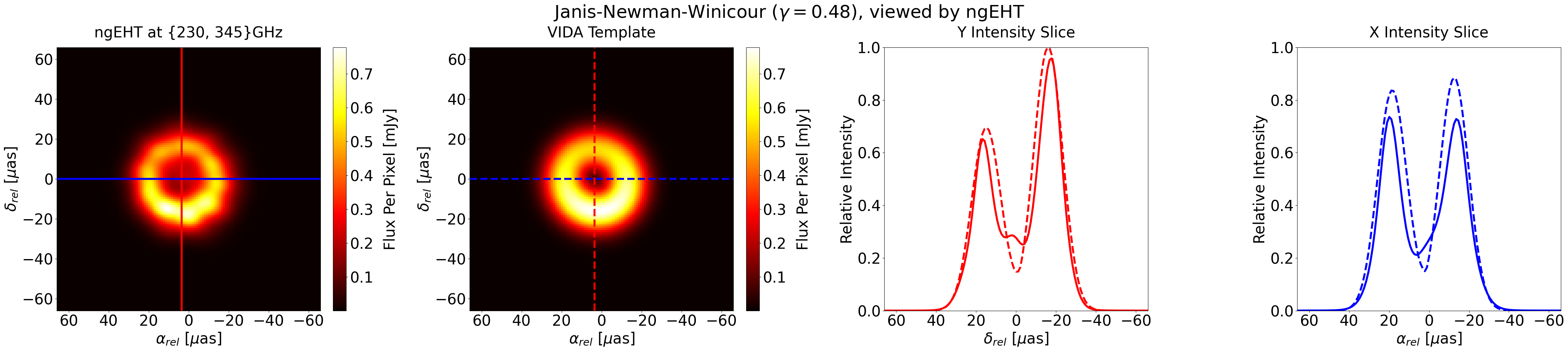}
 \caption{Reconstructed images (left) and VIDA fit with a ring template (right) for the accretion disks as seen by the ngEHT array at both observing frequencies $\nu=\{230,345\}$ GHz. In the right panel we investigate the  variation of the radiation intensity across the horizontal  (blue line) and  vertical (red line) cross-section through the center of the fitted ellipse. The intensity cross-sections are compared for the reconstructed images (solid lines) and the fitted VIDA templates (dashed lines). }
\label{fig:VIDA_ngEHT_1}
\end{figure}

\begin{table}[h!]
\centering
\begin{tabular}{c|c|c|c|c}
            \hline
		{template parameter} & {Schw.}&{\thead{Kerr \\ (a=0.5)}}&{Gauss-Bonnet}&{JNW}
		\\\hline\hline
		$\sigma$ {(ring width [$\mu$as])} & 5.97& 5.98& 6.95& 7.69
		\\
		$\tau$ {(ellipticity)} & 0.04& 0.04& 0.05& 0.06
		\\
		$\xi_\tau$ {(ellipse orientation)} & -1.93& 1.24& -1.95& -1.96
		\\
		$s$ {(slash)} & 0.28& 0.34& 0.28& 0.16
		\\
		$\xi_s$ {(slash orientation)} & -1.79& -1.83& -1.74& -1.76
		\\\hline
		$r_0$ {(ring radius [$\mu$as])} & 21.2& 21.2& 21.1&15.4
		\\
		$x_0$ {(offset RA [$\mu$as])} & -8.35& -5.89& -11.47& 1.97
		\\
		$y_0$ {(offset DEC [$\mu$as])} & 11.85& 10.94& 9.32& 1.79
		\\\hline\hline
		{  optimized divergence} & 0.002& 0.003& 0.004& 0.002
            \\ \hline
	\end{tabular}
 \caption{Best-fit parameters of the VIDA ring template for the accretion disk images as seen by the 2022 EHT array. }
 \label{table:VIDA_2022}
\end{table}

\bigskip

\begin{table}[h!]
\centering
\begin{tabular}{c|c|c|c|c}
            \hline
		{template parameter} & {Schw.}&{\thead{Kerr \\ (a=0.5)}}&{Gauss-Bonnet}&{JNW}
		\\\hline\hline
		$\sigma$ {(ring width [$\mu$as])} & 5.80&5.81&6.76&7.56
		\\
		$\tau$ {(ellipticity)} & 0.04&0.04&0.05&0.06
		\\
		$\xi_\tau$ {(ellipse orientation)} & -2.34&-2.30&-2.31&-2.45
		\\
		$s$ {(slash)} & 0.23&0.34&0.28&0.16
		\\
		$\xi_s$ {(slash orientation)} & -1.78&-1.83&-1.73&-1.78
		\\\hline
		$r_0$ {(ring radius [$\mu$as])} & 21.2&21.3&21.2&15.6
		\\
		$x_0$ {(offset RA [$\mu$as])} & 0.74&9.09&-0.34&3.14
		\\
		$y_0$ {(offset DEC [$\mu$as])} & 9.65& 14.49&10.84&0.03
		\\\hline\hline
		{  optimized divergence} & 0.003&0.003&0.005&0.002
            \\ \hline
	\end{tabular}
 \caption{Best-fit parameters of the VIDA ring template for the accretion disk images as seen by the ngEHT array at the observing frequency $\nu=230$ GHz. }
 \label{table:VIDA_ngEHT}
\end{table}

\bigskip

\begin{table}[h!]
\centering
\begin{tabular}{c|c|c|c|c}
            \hline
		{template parameter} & {Schw.}&{\thead{Kerr \\ (a=0.5)}}&{Gauss-Bonnet}&{JNW}
		\\\hline\hline
		$\sigma$ {(ring width [$\mu$as])} & 4.05&4.06&5.08&5.97
		\\
		$\tau$ {(ellipticity)} & 0.03&0.04&0.04&0.05
		\\
		$\xi_\tau$ {(ellipse orientation)} & -2.11&-2.12&1.00&-2.21
		\\
		$s$ {(slash)} & 0.36&0.45&0.38&0.23
		\\
		$\xi_s$ {(slash orientation)} & -1.74&-1.78&-1.73&-1.65
		\\\hline
		$r_0$ {(ring radius [$\mu$as])} & 21.0&21.0&21.4&15.6
		\\
		$x_0$ {(offset RA [$\mu$as])} & 0.89&9.30&-0.23&4.22
		\\
		$y_0$ {(offset DEC [$\mu$as])} & 1.00& 14.91&11.36&0.02
		\\\hline\hline
		{  optimized divergence} & 0.009&0.009&0.018&0.010
            \\ \hline
	\end{tabular}
 \caption{Best-fit parameters of the VIDA ring template for the accretion disk images as seen by the ngEHT array at the observing frequency $\nu=345$ GHz. }
 \label{table:VIDA_ngEHT_1}
\end{table}

\bigskip

\begin{table}[h!]
\centering
\begin{tabular}{c|c|c|c|c}
            \hline
		{template parameter} & {Schw.}&{\thead{Kerr \\ (a=0.5)}}&{Gauss-Bonnet}&{JNW}
		\\\hline\hline
		$\sigma$ {(ring width [$\mu$as])} & 5.27&5.28&6.33&7.15
		\\
		$\tau$ {(ellipticity)} & 0.04&0.04&0.04&0.05
		\\
		$\xi_\tau$ {(ellipse orientation)} & 0.88&0.90&-2.25&0.74
		\\
		$s$ {(slash)} & 0.31&0.38&0.31&0.18
		\\
		$\xi_s$ {(slash orientation)} & -1.77&-1.81&-1.73&-1.72
		\\\hline
		$r_0$ {(ring radius [$\mu$as])} & 21.1&21.2&21.2&15.5
		\\
		$x_0$ {(offset RA [$\mu$as])} & 0.79&9.17&-0.31&3.48
		\\
		$y_0$ {(offset DEC [$\mu$as])} & 9.80& 14.66&11.02&0.04
		\\\hline\hline
		{  optimized divergence} & 0.005&0.005&0.008&0.004
            \\ \hline
	\end{tabular}
 \caption{Best-fit parameters of the VIDA ring template for the accretion disk images as seen by the ngEHT array at both observing frequencies $\nu=\{230,345\}$ GHz. }
 \label{table:VIDA_superposition}
\end{table}

\end{document}